\documentstyle[epsf]{mn}

\newcommand{\bell}{\mbox{\boldmath$l$}}
\newcommand{\hatbell}{\mbox{\boldmath$\hat l$}}
\newcommand{\ba}{\mbox{\boldmath$a$}}
\newcommand{\be}{\mbox{\boldmath$e$}}
\newcommand{\bm}{\mbox{\boldmath$m$}}
\newcommand{\bn}{\mbox{\boldmath$n$}}
\newcommand{\hatbn}{\mbox{\boldmath$\hat n$}}
\newcommand{\br}{\mbox{\boldmath$r$}}
\newcommand{\bv}{\mbox{\boldmath$v$}}
\newcommand{\bF}{\mbox{\boldmath$F$}}
\newcommand{\bG}{\mbox{\boldmath$G$}}
\newcommand{\bL}{\mbox{\boldmath$L$}}
\newcommand{\bS}{\mbox{\boldmath$S$}}
\newcommand{\bT}{\mbox{\boldmath$T$}}
\newcommand{\sI}{{\cal I}}

\title{Precessing warped accretion discs in X-ray binaries}

\author[G. I. Ogilvie and G. Dubus]
  {G. I. Ogilvie$^{1,2}$
  and G. Dubus$^{3}$\\
  $^1$Max-Planck-Institut f\"ur Astrophysik,
  Karl-Schwarzschild-Stra\ss e 1, Postfach 1523,
  D-85740 Garching bei M\"unchen, Germany\\
  $^2$Institute of Astronomy, University of Cambridge, Madingley Road,
  Cambridge CB3 0HA\\
  $^3$Astronomical Institute `Anton Pannekoek', University of
  Amsterdam, Kruislaan 403, 1098 SJ Amsterdam, The Netherlands}

\begin{document}

\maketitle

\label{firstpage}

\begin{abstract}
  We study the radiation-driven warping of accretion discs in the
  context of X-ray binaries.  The latest evolutionary equations are
  adopted, which extend the classical alpha theory to time-dependent
  thin discs with non-linear warps.  We also develop accurate,
  analytical expressions for the tidal torque and the radiation
  torque, including self-shadowing.
  
  We investigate the possible non-linear dynamics of the system within
  the framework of bifurcation theory.  First, we re-examine the
  stability of an initially flat disc to the Pringle instability.
  Then we compute directly the branches of non-linear solutions
  representing steadily precessing discs.  Finally, we determine the
  stability of the non-linear solutions.  Each problem involves only
  ordinary differential equations, allowing a rapid, accurate and well
  resolved solution.
  
  We find that radiation-driven warping is probably not a common
  occurrence in low-mass X-ray binaries.  We also find that stable,
  steadily precessing discs exist for a narrow range of parameters
  close to the stability limit.  This could explain why so few systems
  show clear, repeatable `super-orbital' variations.  The best
  examples of such systems, Her X-1, SS 433 and LMC X-4, all lie close
  to the stability limit for a reasonable choice of parameters.
  Systems far from the stability limit, including Cyg X-2, Cen X-3 and
  SMC X-1, probably experience quasi-periodic or chaotic variability
  as first noticed by Wijers \& Pringle.  We show that
  radiation-driven warping provides a coherent and persuasive
  framework but that it does not provide a generic explanation for the
  long-term variabilities in all X-ray binaries.
\end{abstract}

\begin{keywords}
  accretion, accretion discs -- binaries: close -- hydrodynamics --
  instabilities -- X-rays: stars.
\end{keywords}

\section{Introduction}

A growing number of X-ray binaries are known to exhibit more or less
periodic variability on time-scales much longer than their orbital
periods (see Priedhorsky \& Holt 1987; Smale \& Lochner 1992; White,
Nagase \& Parmar 1995; Corbet \& Peele 1997).  The clearest examples
of such `super-orbital' periods can be found in Her X-1, SS 433 and
LMC X-4.  They have been known for a long time and are well
documented.  Different explanations have been proposed for the long
periods of these systems but there seems to be a converging opinion,
from both the observational and theoretical viewpoints, that these
periods represent the precession of a tilted accretion disc (Katz
1973; Petterson 1975; Schwarzenberg-Czerny 1992; Wijers \& Pringle
1999 and references therein).  As well as periodically obscuring the
X-ray source, a tilted disc casts an asymmetric shadow on the
irradiated companion star which well reproduces the optical
light-curves (e.g. Gerend \& Boynton 1976; Leibowitz 1984; Heemskerk
\& van Paradijs 1989; Still et~al. 1997).

It has long been recognized that the precession of a tilted disc could
be caused by the tidal force of the companion star (Katz 1973).
Consider the disc to be composed of a sequence of concentric circular
rings, tilted with respect to the binary orbital plane.  The rings
behave gyroscopically owing to their rapid rotation.  If they did not
interact with each other, the tidal torque would cause each ring to
precess retrogradely, about an axis perpendicular to the binary plane,
at a rate that depends on the radius of the ring, resulting in a rapid
twisting of the disc.  However, a fluid disc tends to resist this by
establishing an internal torque between neighbouring rings.  This can
be arranged so that the net torque on each ring is such as to produce
a single, uniform precession rate, allowing the disc to precess
coherently.  However, the internal torque can only be established if
the disc becomes warped.  This is accompanied by dissipation of energy
and the disc settles into the binary plane (Lubow \& Ogilvie 2000;
Bate et~al. 2000).

A mechanism is therefore required to excite the tilt continuously.
Radiation-driven warping has been identified as a plausible way to do
this.  If the outer part of the disc is warped, it intercepts
radiation from the central point source.  If this radiation is
absorbed and re-emitted parallel to the local normal to the disc
surface, the disc experiences a torque from the radiation pressure
(Petterson 1977; Pringle 1996).  The albedo of the disc is probably
irrelevant since scattered photons leaving the disc will also have a
distribution of momenta peaked around the local normal (see
Appendix~A4).  A small tilt grows exponentially if it is twisted such
that the line of nodes forms a leading spiral, and if the luminosity
is sufficient to overcome dissipative processes that tend to flatten
the disc (Pringle 1996).

Such a mechanism would apply very well to X-ray binaries where the
optical emission of the accretion disc is dominated by X-ray
reprocessing (van Paradijs \& McClintock 1994).  This does not
preclude other driving mechanisms for warps, in particular wind
torques (Schandl \& Meyer 1994), but radiation-driven warping does
have the advantage of relying on well-defined physics.

The first question to be addressed is whether an initially flat disc
is unstable to radiation-driven warping.  Previous studies (Pringle
1996; Maloney, Begelman \& Pringle 1996; Maloney \& Begelman 1997;
Maloney, Begelman \& Nowak 1998) concluded that discs in X-ray
binaries are indeed likely to be unstable.  Although instructive,
these studies neglected self-shadowing of the disc and were not
satisfying in their treatment of the boundary conditions, which can be
crucial to this problem.  Furthermore, there was some uncertainty as
to what equations to use.  Recent progress (Ogilvie 1999, 2000) has
led to a practical set of equations describing non-linear warps in
thin discs.  Although only one-dimensional, the equations are derived
directly from the basic fluid-dynamical equations in three dimensions,
within the context of the alpha theory.

The second question concerns the non-linear development of the
instability.  In particular, does it lead to steadily precessing discs
as suggested by observations of some systems?  Recently, Wijers \&
Pringle (1999) investigated numerically the non-linear evolution of
the instability and found a complex variety of behaviour.  Their study
was based on a direct integration of the earlier evolutionary
equations of Pringle (1992).  The calculations were restricted to low
resolution owing mainly to the laborious treatment of self-shadowing.

In this paper, we present a complementary study to that of Wijers \&
Pringle (1999).  In Section~2 we present our formulation of the
dynamical problem, based on the new evolutionary equations of Ogilvie
(2000).  We also derive accurate, analytical expressions for the tidal
torque and the radiation torque, including self-shadowing (see the
Appendices).  In Section~3 we re-examine the stability of an initially
flat disc to radiation-driven warping.  We then search directly for
non-linear solutions representing steadily precessing discs and
determine their stability (Section~4).  All this can be achieved with
high precision and resolution because it involves solving only
ordinary differential equations (ODEs).  These solutions would be
directly applicable to systems such as Her X-1.  Finally, in Section~5
we compare our results with observations.  For the most part, we
investigate generic properties in this paper, and leave the detailed
modelling of individual systems to future work.

\section{Dynamics of warped discs}

\subsection{Background}

In general, the dynamics of a warped accretion disc in an X-ray binary
is an extremely difficult problem of radiation magnetohydrodynamics in
three dimensions.  Furthermore, the evolution must be followed for
many times longer than the characteristic orbital time-scale of the
disc if one is to account for the observed behaviour.  These
considerations mean that an {\it ab initio\/} direct numerical
treatment of the problem is completely unfeasible at present.

Considerable progress has been made in recent years towards developing
a reduced description of warped accretion discs that is only
one-dimensional and follows the evolution on the long, viscous
time-scale.  This approach was introduced by Papaloizou \& Pringle
(1983) and developed by Pringle (1992).  It is based on writing down
plausible conservation equations in one dimension for a disc composed
of a continuum of thin circular rings that are in centrifugal balance
but have varying inclinations.  The rings interact by means of viscous
torques.

Although this approach is intentionally simplistic in origin
(Papaloizou \& Pringle 1983), it has been shown that the same form of
equations, with minor modifications, can be derived formally from the
Navier-Stokes equation in three dimensions (Ogilvie 1999).  This
derivation also provides the non-trivial relation between the
effective viscosity coefficients appearing in the reduced equations
and the small-scale effective viscosity that acts on motions with a
length-scale comparable to the disc thickness.  The resulting theory
can then be seen as a non-linear extension of the existing linear
theory of bending waves in viscous Keplerian discs (Papaloizou \&
Pringle 1983).

To make this derivation possible, of course, certain assumptions must
be made.  The first is that the disc is thin, which is expected to be
satisfied under a wide range of conditions.  Indeed, the resulting
theory is formally an asymptotic theory for thin discs.  The second is
that the agent responsible for angular momentum transport in accretion
discs can be treated as an isotropic effective viscosity.  Although it
is now widely believed that this agent is magnetohydrodynamic
turbulence resulting from the magnetorotational instability (Balbus \&
Hawley 1998), our understanding of how the turbulence responds to
imposed motions other than a Keplerian shear flow is extremely limited
(Abramowicz, Brandenburg \& Lasota 1996).  A first attempt to measure
this response for motions of the type induced in a warped disc
(Torkelsson et~al. 2000) suggests that the assumption of an isotropic
effective viscosity may be reasonable as a first approximation, and,
in any case, no better model has been proposed.  The third assumption
is that the resonant bending-wave propagation found in inviscid,
Keplerian discs may be neglected.  This condition is satisfied if the
viscosity parameter exceeds the angular semi-thickness of the disc
($\alpha\ga H/r$).\footnote{Even if $\alpha\la H/r$, resonant wave
  propagation is likely to be compromised by non-linear effects,
  including a parametric instability (Gammie, Goodman \& Ogilvie
  2000).} Further assumptions are of a more technical nature, but
include the assumption that the fluid is polytropic.

In a recent paper (Ogilvie 2000) this method was extended to include
non-trivial thermodynamics.  Instead of assuming the fluid to be
polytropic, one now accounts fully for viscous dissipation and
radiative energy transport within the Rosseland approximation.  The
fluid is treated as an ideal gas, and a power-law opacity function,
such as Thomson or Kramers opacity, is assumed.  The resulting theory
forms a closed system and may be considered as the logical extension
of the alpha-disc theory (Shakura \& Sunyaev 1973) to time-dependent,
warped accretion discs.  In the absence of a warp, it reduces exactly
to the usual alpha theory, except that the vertical structure of the
disc is computed accurately by solving the differential equations,
rather than by the usual algebraic approximations.

\subsection{Basic equations}

We consider a binary system with a compact star (or black hole) of
mass $M_1$, about which the disc orbits, and a companion star of mass
$M_2$, in a circular orbit of radius $r_{\rm b}$ and angular frequency
\begin{equation}
  \Omega_{\rm b}=\left[{{G(M_1+M_2)}\over{r_{\rm b}^3}}\right]^{1/2}.
\end{equation}
We adopt Cartesian coordinates $(x,y,z)$ in a non-rotating frame of
reference centred on the compact star, such that the orbit of the
companion lies in the $xy$-plane and has a positive sense.

The dynamical system is described by partial differential equations in
coordinates $(r,t)$, where $r$ is the distance from the central object
and $t$ is the time.  At radius $r$, the disc rotates in centrifugal
balance, with orbital angular velocity $\Omega(r)$ and specific
angular momentum $h(r)=r^2\Omega$.  The state of the disc is
characterized by a surface density $\Sigma(r,t)$ and a tilt vector
$\bell(r,t)$, which is a unit vector parallel to the local orbital
angular momentum of the disc.  The orbital angular momentum density is
then $\bL(r,t)=\Sigma h\,\bell$.  Throughout this paper we take the
angular velocity to be Keplerian, i.e. $\Omega=(GM_1/r^3)^{1/2}$.

The general forms of the conservation equations for mass and angular
momentum are
\begin{equation}
  {{\partial\Sigma}\over{\partial t}}+
  {{1}\over{r}}{{\partial}\over{\partial r}}(rv\Sigma)=S
  \label{dsigma}
\end{equation}
and
\begin{equation}
  {{\partial\bL}\over{\partial t}}+
  {{1}\over{r}}{{\partial}\over{\partial r}}(rv\bL)=
  {{1}\over{r}}{{\partial\bG}\over{\partial r}}+\bT,
  \label{dl}
\end{equation}
respectively.  Here $v(r,t)$ is the mean radial velocity and
$2\pi\bG(r,t)$ is the internal torque in the disc.  The terms $S(r,t)$
and $\bT(r,t)$ represent the sources (per unit area) of mass and
angular momentum, respectively. If necessary, these terms should be
considered as the mean sources averaged over the orbital time-scale.

The total external torque consists of three parts,
\begin{equation}
  \bT=\bT_{\rm add}+\bT_{\rm rad}+\bT_{\rm tide}.
\end{equation}
The first represents the angular momentum added to the disc along with
the accretion stream.  The radiation torque $\bT_{\rm rad}$, including
the effects of self-shadowing, is evaluated in Appendix~A, where we
find (cf. equation \ref{trad})
\begin{equation}
  \bT_{\rm rad}=-{{L_\star}\over{12\pi rc}}\,\bell\times\ba.
\end{equation}
Here $L_\star$ is the luminosity of the central source (assumed
isotropic), $c$ the speed of light and $\ba$ a dimensionless vector
depending on the shape of the disc.  The tidal torque $\bT_{\rm
  tide}$, averaged over the binary orbit, is evaluated in Appendix~B,
where we find (cf. equation \ref{ttide})
\begin{equation}
  \bT_{\rm tide}=-{{3GM_2\Sigma r^2}\over{4r_{\rm b}^3}}(\be_z\cdot\bell)
  (\be_z\times\bell)\left(1+\cdots\right).
\end{equation}
The final factor (not shown explicitly here) represents an advance
over previous work; to omit it typically results in an error of $15\%$
at $r/r_{\rm b}=0.3$.  Any other potential torques, such as
self-gravitational, magnetic or gravitomagnetic torques, are
neglected.  Note that the acceleration of the origin of the coordinate
system does not contribute a torque (Wijers \& Pringle 1999).

\subsubsection{Internal torque}

Detailed analysis of the internal hydrodynamics of a warped disc
(Ogilvie 1999, 2000) results in an expression for the internal torque
of the form
\begin{equation}
  \bG=\sI r^2\Omega^2
  \left(Q_1\,\bell+Q_2r{{\partial\bell}\over{\partial r}}+
  Q_3r\,\bell\times{{\partial\bell}\over{\partial r}}\right),
  \label{G}
\end{equation}
where
\begin{equation}
  \sI=\Big\langle\int\rho z^2\,{\rm d}z\Big\rangle
\end{equation}
is the second vertical moment of the density, and $Q_1$, $Q_2$ and
$Q_3$ are dimensionless coefficients.  (In this definition only, $z$
denotes the distance from the mid-plane of the warped disc, and the
angle brackets denote azimuthal averaging.)  The generalized alpha
theory (Ogilvie 2000) provides a relation between $\sI$ and $\Sigma$
of the form
\begin{equation}
  \sI=Q_5C_\sI\alpha^{1/7}\Sigma^{10/7}\Omega^{-12/7},
  \label{I}
\end{equation}
where $Q_5$ is a further dimensionless coefficient, equal to unity for
a flat disc, $C_\sI$ is a constant, and $\alpha$ is the dimensionless
viscosity parameter.\footnote{This equation is equivalent to the usual
  relation between the integrated viscous stress `$\nu\Sigma$' and the
  surface density $\Sigma$ in a flat disc.  The coefficient $Q_5$
  reflects the thickening of the disc in response to increased viscous
  dissipation associated with motions induced by the warp.} It is
assumed here that the disc is Keplerian and gas-pressure-dominated,
and that the opacity law is of the Kramers form $\kappa\propto\rho
T^{-7/2}$.  The value of the constant $C_\sI$ is approximately
  $4.4\times10^{12}$ in CGS units (${\rm cm}^{20/7}{\rm g}^{-3/7}{\rm
    s}^{-12/7}$) (Ogilvie 2000). Although in the inner parts of the
disc Thomson opacity may in fact dominate, and radiation pressure may
be important, it will be seen that these inner parts play only a
passive role in the dynamics, and equation (\ref{I}) is sufficient for
our purposes.

The dimensionless coefficients $Q_i$ depend, most importantly, on the
shear viscosity parameter, $\alpha$, and on the dimensionless
amplitude of the warp,
\begin{equation}
  |\psi|=\left|{{\partial\bell}\over{\partial\ln r}}\right|.
\end{equation}
There are much less significant dependences on the adiabatic exponent
$\gamma$ and the bulk viscosity parameter $\alpha_{\rm b}$.  We assume
that $\gamma$, $\alpha$ and $\alpha_{\rm b}$ are constants, and set
$\gamma=5/3$ and $\alpha_{\rm b}=0$ throughout.

In this formulation, the internal torque is derived consistently from
the basic fluid-dynamical equations in three dimensions, within the
context of the alpha theory (i.e. the assumption that the effective
dynamic viscosity is proportional to the local pressure).  The torque
responds dynamically to changes in $\Sigma$ and increases in response
to the enhanced dissipation that results from a strong warp.  This is
different from the formulation of Wijers \& Pringle (1999), who
assumed the kinematic viscosity $\nu$ to be a fixed function of
radius, independent of $\Sigma$ and $|\psi|$.

For small-amplitude warps, the coefficients have expansions
\begin{equation}
  Q_i=Q_{i0}+O(|\psi|^2),
  \label{expandq}
\end{equation}
where
\begin{eqnarray}
  Q_{10}=-{{3\alpha}\over{2}},\qquad
  &&Q_{20}={{1+7\alpha^2}\over{\alpha(4+\alpha^2)}},\nonumber\\
  Q_{30}={{3(1-2\alpha^2)}\over{2(4+\alpha^2)}},
  &&Q_{50}=1.
\end{eqnarray}
We will consider first the case of {\it linear hydrodynamics}, in
which the $Q_i$ are simply given the values $Q_{i0}$, independent of
$|\psi|$.  Note that, even under this assumption, equations
(\ref{dsigma}) and (\ref{dl}) do not form a linear system.  We will
also consider {\it non-linear hydrodynamics}, in which the dependences
of the $Q_i$ on $|\psi|$ (and on $\gamma$ and $\alpha_{\rm b}$) are
fully taken into account by solving a certain system of ODEs (Ogilvie
2000).  Since only $|\psi|$ varies within a given disc model, one can
tabulate the functions $Q_i(|\psi|)$ beforehand, and simply
interpolate during the course of the calculation.  In our calculations
the amplitude never exceeded $|\psi|=1$.

\subsubsection{Mass input}

In the case of a disc in a low-mass X-ray binary, the mass source $S$
is associated with the accretion stream that originates at the $L_1$
point on the surface of the Roche-lobe-filling companion star.  When
the disc is warped, the stream will intersect it at a point that
depends on the instantaneous shape of the disc and on the binary phase
(Schandl 1996).  In general, the stream reaches in to the vicinity of
its circularization radius $r_{\rm c}$, which is much smaller than the
outer radius $r_{\rm o}$ of the disc.  Although it is feasible to
model this interaction to some extent within the framework of
equations (\ref{dsigma}) and (\ref{dl}), for simplicity, we follow
Wijers \& Pringle (1999) in assuming that the mass input is localized
at a single radius which may be identified approximately as $r_{\rm
  c}$.  The added mass has the same specific angular momentum as a
circular Keplerian orbit of radius $r_{\rm c}$ and aligned with the
binary plane.  We then have
\begin{eqnarray}
  S&=&{{\dot M}\over{2\pi r}}\,\delta(r-r_{\rm c}),\\
  \bT_{\rm add}&=&{{\dot Mh}\over{2\pi r}}\,\delta(r-r_{\rm c})\,\be_z,
\end{eqnarray}
where $\dot M$ is the steady rate of mass supply.  We defer a more
accurate treatment to future work.

\subsubsection{Boundary conditions}

At the inner radius $r=r_{\rm i}$ of the disc, there are several
different possible situations, depending on the nature of the central
object.  In the case of a black hole, the disc is likely to end
(effectively) near the marginally stable circular orbit, where it
makes a transition to a rapidly plunging spiral flow.  At such a
point, the surface density and internal torque fall essentially to
zero (although see Krolik 1999 for an alternative viewpoint).  This
possibility may also occur in the case of a weakly magnetized neutron
star; but if not, the disc is likely to extend up to a boundary layer
on the stellar surface.  Finally, in the case of a strongly magnetized
neutron star, the disc is likely to be truncated at a magnetospheric
boundary.  In this paper we apply the simple boundary conditions
$\bG={\bf0}$ and $\partial\bell/\partial r={\bf0}$ at the inner
radius.  (These are independent, since the first already implies
$\sI=0$.)  These conditions state that the internal torque vanishes at
the inner radius, and that the disc is locally flat; however, the
inclination of the inner edge is allowed to vary freely without having
any preferred orientation.\footnote{If the disc is terminated by a
  magnetosphere, a preferred orientation may well be introduced, but
  we neglect this possibility.  In fact, recent analyses of the pulse
  shape changes in Her X-1 (Blum \& Kraus 2000; Scott, Leahy \& Wilson
  2000) do not indicate that the inner disc is forced into alignment
  with the spin axis of the neutron star, as might have been
  expected.}

At the outer radius $r=r_{\rm o}$, the disc is expected to be
truncated by a localized tidal torque associated with the $m=2$
Lindblad resonance (Papaloizou \& Pringle 1977; Paczy\'nski 1977).
The radius of the Lindblad resonance is unchanged by the tilt of the
disc, and Larwood et~al. (1996) have found that tidal truncation is
only marginally affected. We apply the boundary conditions
$\bell\times\bG={\bf0}$ and $v=0$.  These imply that
$\partial\bell/\partial r={\bf0}$, and are appropriate if the
truncating tidal torque is parallel to $\bell$, and if only angular
momentum, rather than mass, is removed through the outer boundary.
Again, these boundary conditions do not specify any preferred
orientation for the outer edge of the disc.

These boundary conditions are intended to be close to those used by
Wijers \& Pringle (1999), which were, however, implemented in an
explicitly grid-dependent manner.

\subsection{Non-dimensionalization}

At least three reasonable choices of units are possible when
implementing the equations numerically.  First, the equations could be
implemented directly in CGS units.  This might be appropriate when
studying a particular system with well determined parameters, and has
the advantage that the results can be compared immediately with the
observed properties.  In practice, however, many critical parameters
are poorly known, even for the intensely studied system Her X-1.
Moreover, this direct approach obscures the scaling relations that
exist between different systems.

Alternatively, a system of units could be adopted that is based on the
geometry of the binary orbit.  The radius and angular frequency of the
binary orbit would be taken as basic units.  This approach would be
appropriate when studying a system in which the properties of the
binary are well constrained, but perhaps the mass accretion rate and
luminosity are less well known.

In this paper we adopt a third system of units which is based on the
accretion rate and the properties of the central object.  This is
particularly suitable for studying the radiation-driven instability in
isolation, and the effect of the tidal torque can be added as an
external perturbation.  This will be important, for example, in order
to demonstrate that the radiation-driven instability can naturally
explain retrograde precession even without the help of the tidal
torque.  This choice of units also has the major advantage of removing
the dependence on the mass accretion rate, one of the most poorly
known parameters, in the dimensionless equations.

We therefore express $r$ (and $r_{\rm i}$, $r_{\rm o}$, $r_{\rm c}$
and $r_{\rm b}$) in units of $GM_1/c^2$ (half the Schwarzschild radius
of the central object), $\Omega$ in units of $c^3/GM_1$, $\bG$ in
units of $(\dot M/2\pi)(GM_1/c)$, $\sI$ in units of $(\dot
M/2\pi)(GM_1/c^3)$, $\Sigma$ in units of $C_\sI^{-7/10}(\dot
M/2\pi)^{7/10}(c^3/GM_1)^{1/2}$ and the modal or precession
frequencies ($\omega$ or $\omega_{\rm p}$, introduced below) in units
of $C_\sI^{7/10}(\dot M/2\pi)^{3/10}(GM_1)^{-3/2}c^{5/2}$.

This simplifies matters such that the angular velocity is
$\Omega=r^{-3/2}$, the factor $C_\sI$ drops out of equation (\ref{I}),
and the radiation torque becomes (cf. equation \ref{trad})
\begin{equation}
  \bT_{\rm rad}=-{{\epsilon}\over{6r}}\,\bell\times\ba,
\end{equation}
where the dimensionless parameter $\epsilon=L_\star/\dot Mc^2$ is the
accretion efficiency.  The tidal torque becomes (cf. equation
\ref{ttide})
\begin{equation}
  \bT_{\rm tide}=-f_{\rm tide}{{3\Sigma r^2}\over{4r_{\rm b}^3}}
  (\be_z\cdot\bell)(\be_z\times\bell)(1+\cdots),
\end{equation}
where
\begin{equation}
  f_{\rm tide}=qC_\sI^{-7/10}(\dot M/2\pi)^{-3/10}(GM_1c)^{1/2}
  \label{ftide}
\end{equation}
is a dimensionless parameter, with $q=M_2/M_1$.

\section{Stability of a steady, flat disc to warping}

The first dynamical problem we consider is the linear stability of an
initially flat disc to warping.  This problem was first considered by
Pringle (1996) within a WKB approximation, and later by Maloney,
Begelman \& Pringle (1996), Maloney \& Begelman (1997) and Maloney,
Begelman \& Nowak (1998), who examined the general properties of
solutions of the linearized equations and only later considered the
effect of an outer boundary condition.  Our investigation differs in
that we consider a disc of given size, compute the discrete complex
frequency eigenvalues of its bending modes, and thus determine whether
it is stable or unstable.  Moreover, our treatment of the
hydrodynamics is based on the new evolutionary equations of Ogilvie
(2000) which are derived directly from the basic fluid-dynamical
equations under the simple hypothesis of an isotropic alpha viscosity.
Finally, we also consider the effect of self-shadowing on the linear
stability problem.  In some previous work this was described as an
intrinsically non-linear effect, but here we are concerned with
perturbations involving tilt angles that exceed the small variations
in $H/r$ across the disc.

\subsection{Solution for a steady, flat disc}

For a steady, flat disc in the binary plane we have
\begin{equation}
  {{\partial\Sigma}\over{\partial t}}=0,\qquad
  {{\partial\bL}\over{\partial t}}={\bf0},\qquad
  \bell=\be_z,\qquad
  \bG=G_z\,\be_z,
\end{equation}
and $\bT_{\rm rad}=\bT_{\rm tide}={\bf0}$.  The mass equation
(\ref{dsigma}) can be integrated to give
\begin{equation}
  rv\Sigma=-{{\dot M}\over{2\pi}}H(r_{\rm c}-r),
\end{equation}
where $H$ is the Heaviside function, and we have noted that $v=0$ at
$r=r_{\rm o}$.  The radial velocity can then be eliminated and we are
left with the angular momentum equation (\ref{dl}) in the form
\begin{equation}
  -{{\dot M}\over{2\pi r}}{{{\rm d}h}\over{{\rm d}r}}H(r_{\rm c}-r)=
  {{1}\over{r}}{{{\rm d}G_z}\over{{\rm d}r}}.
\end{equation}
With $G_z=0$ at $r=r_{\rm i}$, the solution is
\begin{eqnarray}
  G_z&=&-{{\dot M}\over{2\pi}}(h-h_{\rm i}),
  \qquad r_{\rm i}<r<r_{\rm c},\nonumber\\
  G_z&=&-{{\dot M}\over{2\pi}}(h_{\rm c}-h_{\rm i}),
  \qquad r_{\rm c}<r<r_{\rm o},
\end{eqnarray}
where $h_{\rm i}=h(r_{\rm i})$, etc.

From the general expression (\ref{G}) for $\bG$, we have for a flat
disc
\begin{equation}
  G_z=Q_{10}\sI r^2\Omega^2
\end{equation}
and
\begin{equation}
  \sI=C_\sI\alpha^{1/7}\Sigma^{10/7}\Omega^{-12/7}.
\end{equation}
Thus, given the parameters $M_1$, $r_{\rm i}$, $r_{\rm c}$, $r_{\rm
  o}$, $\dot M$, $\alpha$ and $C_\sI$, the quantities $G_z$, $\sI$,
$\Sigma$ and $v$ can be obtained explicitly.

\subsection{Linear perturbations}

Now consider linear Eulerian perturbations of the basic state, denoted
by a prime.  Then
\begin{equation}
  {{\partial\Sigma'}\over{\partial t}}+
  {{1}\over{r}}{{\partial}\over{\partial r}}(rv'\Sigma+rv\Sigma')=0
\end{equation}
and
\begin{equation}
  {{\partial\bL'}\over{\partial t}}+
  {{1}\over{r}}{{\partial}\over{\partial r}}(rv'\bL+rv\bL')=
  {{1}\over{r}}{{\partial\bG'}\over{\partial r}}+\bT',
  \label{dlprimedt}
\end{equation}
where
\begin{equation}
  \bL'=\Sigma'h\,\be_z+\Sigma h\,\bell'.
\end{equation}
It is sufficient to consider the $x$- and $y$-components of equation
(\ref{dlprimedt}), which read
\begin{eqnarray}
  \Sigma h{{\partial\ell_x'}\over{\partial t}}+
  {{1}\over{r}}{{\partial}\over{\partial r}}(rv\Sigma h\ell_x')&=&
  {{1}\over{r}}{{\partial G_x'}\over{\partial r}}+T_x',\nonumber\\
  \Sigma h{{\partial\ell_y'}\over{\partial t}}+
  {{1}\over{r}}{{\partial}\over{\partial r}}(rv\Sigma h\ell_y')&=&
  {{1}\over{r}}{{\partial G_y'}\over{\partial r}}+T_y'.
\end{eqnarray}
Also
\begin{eqnarray}
  G_x'&=&Q_{10}\sI r^2\Omega^2\ell_x'+
  Q_{20}\sI r^3\Omega^2{{\partial\ell_x'}\over{\partial r}}-
  Q_{30}\sI r^3\Omega^2{{\partial\ell_y'}\over{\partial r}},\nonumber\\
  G_y'&=&Q_{10}\sI r^2\Omega^2\ell_y'+
  Q_{20}\sI r^3\Omega^2{{\partial\ell_y'}\over{\partial r}}+
  Q_{30}\sI r^3\Omega^2{{\partial\ell_x'}\over{\partial r}}.
\end{eqnarray}
Introduce a complex notation (Hatchett, Begelman \& Sarazin 1981) in
which
\begin{equation}
  W=\ell_x'+{\rm i}\ell_y',\qquad
  G=G_x'+{\rm i}G_y',\qquad
  T=T_x'+{\rm i}T_y'.
\end{equation}
Then
\begin{eqnarray}
  \lefteqn{\Sigma h{{\partial W}\over{\partial t}}=
  {{1}\over{r}}{{\partial G}\over{\partial r}}+T-
  {{\dot Mh}\over{2\pi r}}W\delta(r-r_{\rm c})}&\nonumber\\
  &&+{{\dot M}\over{2\pi r}}\left(h{{\partial W}\over{\partial r}}+
  {{{\rm d}h}\over{{\rm d}r}}W\right)H(r_{\rm c}-r),
\end{eqnarray}
with
\begin{equation}
  G=Q_{10}\sI r^2\Omega^2W+
  Q_{40}\sI r^3\Omega^2{{\partial W}\over{\partial r}},
\end{equation}
where
\begin{equation}
  Q_{40}=Q_{20}+{\rm i}Q_{30}=
  {{1+2{\rm i}\alpha+6\alpha^2}\over{2\alpha(2+{\rm i}\alpha)}}.
\end{equation}
In this linear problem, only the first terms in the expansions of the
$Q_i$ (equation \ref{expandq}) are required.  There is therefore no
distinction at this stage between the models of linear and non-linear
hydrodynamics defined in Section~2.2.1.

If self-shadowing is neglected, the linearized radiation torque is (from
Appendix~A)
\begin{equation}
  T_{\rm rad}=-\left({{L_\star}\over{12\pi c}}\right)
  {\rm i}{{\partial W}\over{\partial r}}.
\end{equation}
Including self-shadowing, we obtain instead
\begin{eqnarray}
  \lefteqn{T_{\rm rad}=-\left({{L_\star}\over{12\pi c}}\right)
  {\rm i}\left\{
  \left[g_1(\theta_{\rm min}+\pi,0)-g_1(\theta_{\rm max},0)\right]
  \right.}&\nonumber\\
  &&\left.+{\rm i}
  \left[g_2(\theta_{\rm min}+\pi,0)-g_2(\theta_{\rm max},0)\right]
  \right\}{{\partial W}\over{\partial r}},
\end{eqnarray}
with
\begin{eqnarray}
  g_1(\theta,0)&=&{{1}\over{\pi}}(\theta+\cos\theta\sin\theta),\nonumber\\
  g_2(\theta,0)&=&{{1}\over{\pi}}\sin^2\theta.
\end{eqnarray}
The shadow angle $\theta_{\rm s}$ given by equation (\ref{thetas}) can
be expressed in terms of $W$ as
\begin{equation}
  \theta_{\rm s}={{\pi}\over{2}}+\arg(W-\hat W)-
  \arg{{\partial W}\over{\partial r}},
\end{equation}
where $\hat W$ is the tilt variable of the ring that shadows the
  ring under consideration.

The linearized tidal torque is (from Appendix~B)
\begin{eqnarray}
  \lefteqn{T_{\rm tide}=-{{3GM_2\Sigma r^2}\over{4r_{\rm b}^3}}}&
  \nonumber\\
  &&\times\left[1+{{15}\over{8}}\left({{r}\over{r_{\rm b}}}\right)^2+
  {{175}\over{64}}\left({{r}\over{r_{\rm b}}}\right)^4+\cdots\right]
  {\rm i}W.
\end{eqnarray}
This can also be written as
\begin{equation}
  T_{\rm tide}=-{{GM_2\Sigma r}\over{4r_{\rm b}^2}}
  \left[b_{3/2}^{(1)}\left({{r}\over{r_{\rm b}}}\right)\right]
  {\rm i}W,
\end{equation}
where $b_\gamma^{(m)}$ is the Laplace coefficient of celestial mechanics.

We may then seek a normal mode in which
\begin{equation}
  {{\partial W}\over{\partial t}}\mapsto{\rm i}\omega W,
\end{equation}
where $\omega$ is the complex frequency eigenvalue.  The problem then
reduces to a complex linear eigenvalue problem involving a
second-order system of ODEs.  It is convenient to use $W$ and $G$ as
the dependent variables.

The inner radius is a singular point since $\sI=0$ there.  Here we may
apply the arbitrary normalization condition $W=1$.  We also have ${\rm
  d}W/{\rm d}r=0$, $G=0$ and ${\rm d}G/{\rm d}r=-(\dot M/2\pi){\rm
  d}h/{\rm d}r$ for the regular solution.  We then integrate from
$r=r_{\rm i}$ to $r=r_{\rm c}$.

Owing to the presence of a $\delta$-function in the angular momentum
equation, $G$ is discontinuous at $r=r_{\rm c}$, while $W$ is
continuous (Maloney, Begelman \& Nowak 1998).  In a standard notation,
the discontinuity in $G$ is
\begin{equation}
  [G]=\lim_{\delta\to0}
  G\bigg|_{r=r_{\rm c}-\delta}^{r=r_{\rm c}+\delta}=
  {{\dot Mh}\over{2\pi}}W.
\end{equation}
It is trivial to implement this jump condition and integrate further
from $r=r_{\rm c}$ to $r=r_{\rm o}$.

At the outer radius we have the boundary condition ${\rm d}W/{\rm
  d}r=0$.  Thus one complex quantity ($\omega$) must be guessed to
start the integration, and one complex condition must be met at the
outer boundary.  Such a problem is readily solved numerically by
Newton-Raphson iteration.

\subsection{Numerical investigation}

In his initial investigation of the radiation-driven instability,
Pringle (1996) gave a simple approximate criterion for instability of
a disc to warping.  The disc should be unstable when
\begin{equation}
  {{r_{\rm o}}\over{r_{\rm S}}}\ga{{8\pi^2\eta^2}\over{\epsilon^2}},
\end{equation}
where $r_{\rm S}$ is the Schwarzschild radius of the central object
and $\eta=\nu_2/\nu_1$ is the ratio of effective viscosities.  Under
the assumptions of this paper,
\begin{equation}
  \eta=-{{3Q_{20}}\over{Q_{10}}}=
  {{2(1+7\alpha^2)}\over{\alpha^2(4+\alpha^2)}}
  \label{eta}
\end{equation}
in linear theory (see Papaloizou \& Pringle 1983; Ogilvie 1999).  The
approximate criterion can be rewritten for our purposes as
\begin{equation}
  {{r_{\rm b}}\over{GM_1/c^2}}\ga{{16\pi^2\eta^2}\over{\epsilon^2}}
  {{r_{\rm b}}\over{r_{\rm o}}}.
  \label{pringle}
\end{equation}
The critical radius depends strongly on $\epsilon$ and especially on
$\alpha$. Interestingly, it depends on the luminosity of the central
object $L_\star$ only through the accretion efficiency $\epsilon$ and
not through the mass accretion rate.  Based on the properties of a
low-mass X-ray binary with $q\approx1$ (e.g.  Her X-1), we take
$r_{\rm o}/r_{\rm b}=0.3$ (cf. Papaloizou \& Pringle 1977; Paczy\'nski
1977) and $r_{\rm c}/r_{\rm b}=0.09$ (cf. Lubow \& Shu 1975; Flannery
1975).  We also take $r_{\rm i}=6GM_1/c^2$.  The consequences of
varying $q$ and/or of assuming the mass input to occur at the outer
radius are examined in Section~5.1 below.

These considerations lead us to choose $r_{\rm b}$ as the basic
control parameter.  It is also the quantity that varies most
significantly and obviously between different systems.  Two other
important parameters that affect stability are $\alpha$ and
$\epsilon$, which are not well constrained, but it is difficult to
argue why they should vary substantially between different systems.
Our approach, therefore, is to determine numerically the minimum
binary radius for instability as a function of $\alpha$ and $\epsilon$
and to compare this with equation (\ref{pringle}).  We then examine
the effect, if any, of a tidal torque.

In the absence of the radiation and tidal torques, the disc possesses
a discrete set of bending modes with $0$, $1$, $2$, $\dots$ nodes in
the eigenfunction ${\rm Re}(W)$ (say).  These modes are all damped
[${\rm Im}(\omega)>0$] by viscosity and typically precess retrogradely
[${\rm Re}(\omega)<0$] as a result of the internal torque component
with coefficient $Q_3$.  Mode $0$ is exceptional in that it is almost
neutral ($\omega\approx0$) in comparison.  This mode would be the
trivial rigid-tilt mode if the system possessed perfect spherical
symmetry.  However, because the mass input occurs in a definite plane,
the full rotational symmetry of the system is broken and mode $0$
becomes non-trivial.

When the radiation torque is gradually introduced (e.g. by increasing
$\epsilon$ continuously from $0$) the frequency eigenvalues $\omega$
move and the number of nodes in the eigenfunctions may change.  We do
not attempt a proper classification of the modes in the general case,
but continue to label them by continuity with the simple limit of zero
external torque.

In principle, to determine stability, one must check the imaginary
parts of all the frequency eigenvalues of the disc.  In practice, the
higher-order modes can be neglected because they are strongly damped
by viscosity.  The eigenvalues are usually ordered such that mode $0$
becomes unstable first as $r_{\rm b}$ is increased, and the precession
is retrograde.  Under some circumstances, as described below and also
in Section~5.1, mode $1$ can become unstable first, and the precession
is then prograde.

It is also possible to simplify the procedure by solving directly for
a disc possessing a marginally stable mode.  This is done by
restricting $\omega$ to be real but treating $r_{\rm b}$ as a second
eigenvalue.  One then has two real eigenvalues instead of one complex
eigenvalue, and Newton-Raphson iteration can be applied in two real
dimensions.

The results are shown in Figs~1 and 2.  Fig.~1 shows the variation of
the minimum binary radius for instability with $\alpha$ and
$\epsilon$, in the absence of a tidal torque.  It can be seen that the
effect of self-shadowing is to increase the critical radius by about a
factor of $2$.  The approximate criterion (\ref{pringle}) reproduces
quite accurately the steep scalings with $\alpha$ and $\epsilon$, but
systematically overestimates the critical radius.  The reason for this
is probably that Pringle (1996) estimated the wavelength of the
unstable mode as $\la r_{\rm o}$, whereas in fact the effective
wavelength of the modified rigid-tilt mode is somewhat longer than
this. More accurate estimates of the linear stability criterion
  were made by Wijers \& Pringle (1999).

Fig.~2 shows the effect of a tidal torque on the linear stability.
The torque acts against the instability, making the critical radius
larger.  Note that, with a significant tidal torque, mode $1$ becomes
unstable first when $\alpha$ or $\epsilon$ is large, and this results
in prograde precession.

\begin{figure*}
  \centerline{\epsfbox{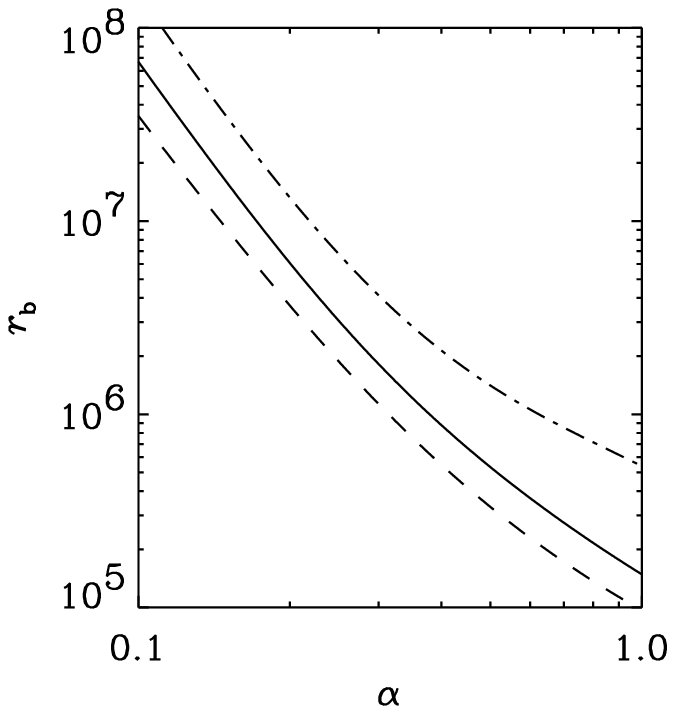}\qquad\epsfbox{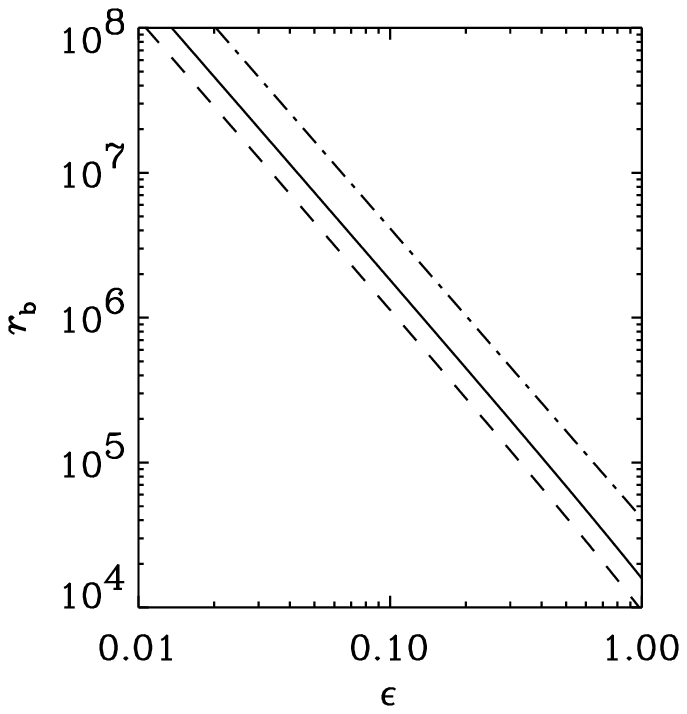}}
  \caption{Critical binary radius for linear instability of a flat disc
    to warping, in units of $GM_1/c^2$.  Left: variation with $\alpha$
    at fixed $\epsilon=0.1$.  Right: variation with $\epsilon$ at
    fixed $\alpha=0.3$.  Dashed line: numerical calculation neglecting
    self-shadowing.  Solid line: numerical calculation including
    self-shadowing.  Dot-dashed line: approximate formula
    (\ref{pringle}).  The tidal torque is not included.}
\end{figure*}

\begin{figure*}
  \centerline{\epsfbox{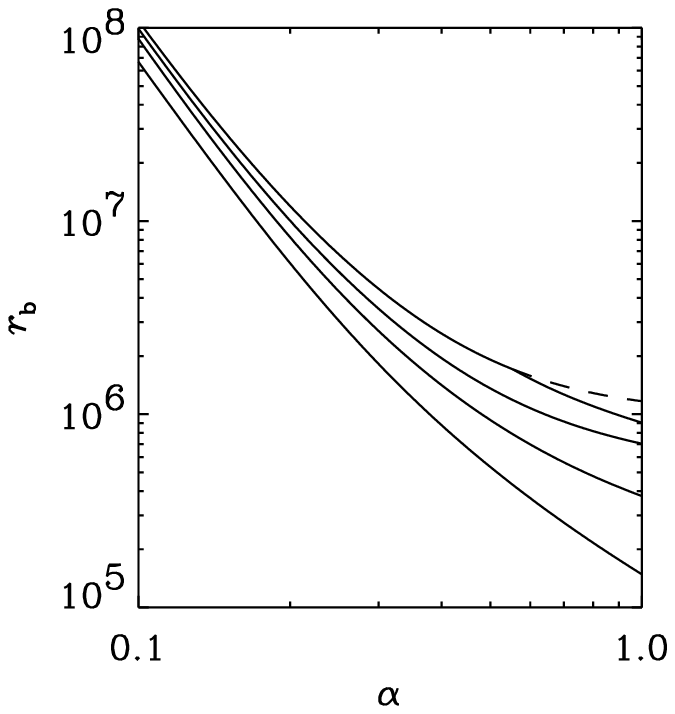}\qquad\epsfbox{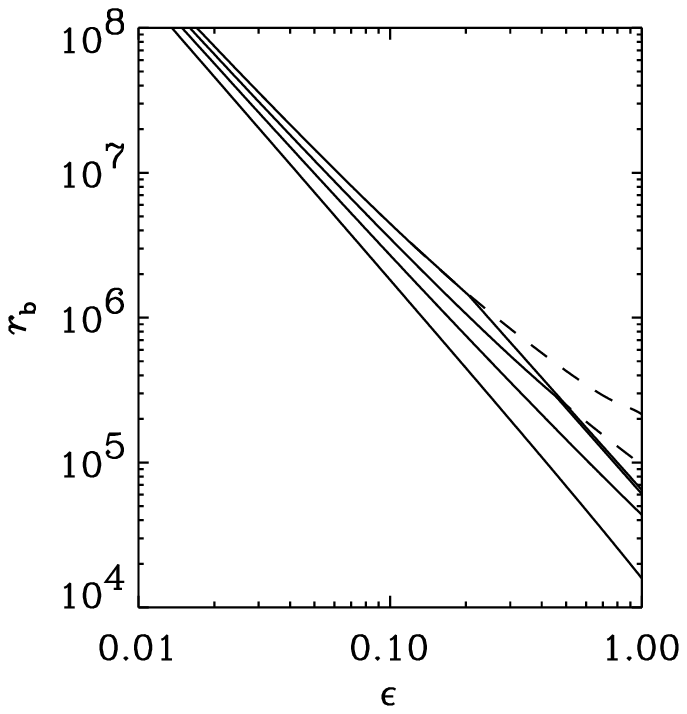}}
  \caption{Effect of a tidal torque on the critical binary radius.  The
    lowest curve in each panel corresponds to the solid line in the
    corresponding panel in Fig.~1.  The other curves show the effect
    of an increasing tidal torque, with (from bottom to top) $f_{\rm
      tide}=10^4$, $2\times10^4$, $3\times10^4$.  In some cases mode
    $1$ (prograde) becomes unstable before mode $0$, leading to a kink
    in the curve.  The dashed segment is then the continuation of the
    marginal stability curve for mode $0$.  A typical value of $f_{\rm
      tide}\approx2.3\times10^4$ can be estimated, based on equation
    (\ref{ftide}) with $q\approx1$,
    $C_\sI\approx4.4\times10^{12}\,{\rm cm}^{20/7}{\rm g}^{-3/7}{\rm
      s}^{-12/7}$, $\dot M\approx10^{18}\,{\rm g}\,{\rm s}^{-1}$ and
    $M_1\approx1.4\,M_\odot$.}
\end{figure*}

\section{Steadily precessing discs}

Having determined the conditions for instability of a flat disc, we
now consider the non-linear solutions that result.  In general these
solutions may be arbitrarily complex and can be obtained only by a
direct integration of the time-dependent equations.  However, we can
use the principles of bifurcation theory (e.g. Iooss \& Joseph 1980)
to try to anticipate and categorize this behaviour.

When the control parameter $r_{\rm b}$ is increased through the
critical value $r_{\rm crit}$ for linear instability, a branch of
non-linear solutions bifurcates from the solution representing a flat
disc.  Owing to the symmetries of the problem, these non-linear
solutions are steady in a rotating frame of reference, although the
angular velocity of that frame remains to be determined as an
eigenvalue.  These solutions therefore represent steadily precessing
discs, which are the simplest possible non-linear outcome of the
instability.

Close to the point of bifurcation, these solutions are very little
warped and nearly obey the linearized equations.  The shape of these
solutions is predicted by the eigenfunction of the marginally stable
mode of linear theory, and the precession rate of the disc corresponds
to the frequency of the marginally stable mode.

The branch of non-linear solutions may extend initially either into
$r_{\rm b}>r_{\rm crit}$ (a supercritical bifurcation) or into $r_{\rm
  b}<r_{\rm crit}$ (a subcritical bifurcation).  It may then turn
around, undergo further bifurcations to more complex solutions, or
simply terminate (since the equations being solved are not globally
analytic).  This behaviour depends on the non-linear terms in the
governing equations.

Our approach is to solve directly for the non-linear solutions
representing steadily precessing discs.  Not only are these the
simplest possible outcome of the instability, they are also
appropriate for explaining those observed systems that display a
regular precession.  They can be found by solving a non-linear
second-order ODE eigenvalue problem, much more quickly and accurately
than by using a time-dependent method.  Both stable and unstable
solutions can be found, and the network of branches of solutions
provides a framework for understanding the more complex non-linear
behaviour that can result.

\subsection{Analysis}

For a steady disc precessing about the binary axis with angular
frequency $\omega_{\rm p}$, we may set
\begin{equation}
  {{\partial\Sigma}\over{\partial t}}=0,\qquad
  {{\partial\bL}\over{\partial t}}=\omega_{\rm p}\,\be_z\times\bL.
\end{equation}
The problem is then reduced to solving a set of ODEs with $\omega_{\rm
  p}$ as an eigenvalue to be determined.

The mass equation (\ref{dsigma}) may be integrated as before to give
\begin{equation}
  rv\Sigma=-{{\dot M}\over{2\pi}}H(r_{\rm c}-r).
\end{equation}
The angular momentum equation (\ref{dl}) then becomes
\begin{eqnarray}
  \lefteqn{\omega_{\rm p}\,\be_z\times\bL=
  {{1}\over{r}}{{{\rm d}\bG}\over{{\rm d}r}}+\bT+
  {{\dot Mh}\over{2\pi r}}\delta(r-r_{\rm c})(\be_z-\bell)}&\nonumber\\
  &&+{{\dot M}\over{2\pi r}}{{{\rm d}(h\bell)}\over{{\rm d}r}}
  H(r_{\rm c}-r).
\end{eqnarray}

It is most natural, when integrating the angular momentum equation, to
use $\bell$ and $\bG$ as the dependent variables.\footnote{In fact the
  three components of $\bell$ are not independent since it is a unit
  vector.  However, the condition $|\bell|=1$ may be used as a check
  on the accuracy of the numerical method.  This condition was
  satisfied to within $10^{-10}$ for all solutions obtained.} In order
to implement the equations numerically, one must be able to determine
both $\Sigma$ and ${\rm d}\bell/{\rm d}r$ from $\bell$ and $\bG$.  One
may proceed by writing
\begin{equation}
  r{{{\rm d}\bell}\over{{\rm d}r}}=|\psi|\,\bm,
\end{equation}
where $\bm$ is a unit vector satisfying $\bell\cdot\bm=0$.  Let
$\bn=\bell\times\bm$, so that $(\bell,\bm,\bn)$ is a right-handed
orthonormal triad.  Then
\begin{equation}
  \bG=\sI r^2\Omega^2\left(Q_1\,\bell+Q_2|\psi|\,\bm+
  Q_3|\psi|\,\bn\right).
\end{equation}
Now construct the vector
\begin{equation}
  \bv={{\bell\times\bG}\over{\bell\cdot\bG}}=
  {{Q_2}\over{Q_1}}|\psi|\,\bn-{{Q_3}\over{Q_1}}|\psi|\,\bm.
\end{equation}
Then
\begin{equation}
  |\bv|^2=\left({{Q_2^2+Q_3^2}\over{Q_1^2}}\right)|\psi|^2
  \label{vsq}
\end{equation}
is an equation to be solved for $|\psi|$.  In the approximation of
linear hydrodynamics this is trivial.  In the non-linear model it is
not so, and one must be aware of the possibility of multiple roots, or
of no root.\footnote{Calculations suggest that the right-hand side of
  equation (\ref{vsq}) is a monotonically increasing function of
  $|\psi|$ under most circumstances, and therefore its inverse is
  unique.  An exception arises when $\alpha$ is small
  ($\alpha\la0.05$), since then $Q_1$ changes from negative to
  positive as $|\psi|$ increases through the value $|\psi|=|\psi|_{\rm
    crit}\approx\sqrt{24}\alpha$.  In that case the inverse function
  is double-valued, having one value less than $|\psi|_{\rm crit}$ and
  one value greater.  One can immediately distinguish between these
  possibilities, since the condition $\sI>0$ implies that ${\rm
    sgn}(Q_1)={\rm sgn}(\bell\cdot\bG)$.  However, this may not be
  relevant because the disc will typically become unstable to
  perturbations under these circumstances (Ogilvie 2000).}

Having obtained $|\psi|$, one determines the coefficients $Q_i$.  Then
\begin{equation}
  \sI={{\bell\cdot\bG}\over{Q_1r^2\Omega^2}},
\end{equation}
in conjunction with equation (\ref{I}), yields $\Sigma$.  Now
\begin{equation}
  \bell\times\bv=-{{Q_2}\over{Q_1}}|\psi|\,\bm-
  {{Q_3}\over{Q_1}}|\psi|\,\bn,
\end{equation}
which allows $\bn$ to be eliminated, giving
\begin{equation}
  {{{\rm d}\bell}\over{{\rm d}r}}=-{{1}\over{r}}
  \left({{Q_1}\over{Q_2^2+Q_3^2}}\right)
  \left(Q_3\,\bv+Q_2\,\bell\times\bv\right),
\end{equation}
as required.

At the inner radius of the disc, owing to the axisymmetry of the
problem, we may set $\ell_y=0$ there without loss of generality, and
then $\bell=(\sin\beta,0,\cos\beta)$, where $\beta$ is the inclination
of the inner disc.  Therefore two real quantities, $\omega_{\rm p}$
and $\beta$, must be guessed when starting the integration.  We then
integrate the angular momentum equation from $r=r_{\rm i}$ to
$r=r_{\rm c}$.

Both the surface density and the derivative of the tilt vector are
discontinuous at $r=r_{\rm c}$, where
\begin{equation}
  [\bG]={{\dot Mh}\over{2\pi}}(\bell-\be_z),
\end{equation}
while $\bell$ itself is continuous.

At the outer radius, we apply the boundary condition
$\bell\times\bG={\bf0}$.  This consists of two independent constraints
on the solution (e.g. the $x$- and $y$- components of the equation
$\bell\times\bG={\bf0}$, provided that the outer edge is not inclined
at exactly $90^\circ$), which are sufficient to determine the two
unknown quantities $\omega_{\rm p}$ and $\beta$ by Newton-Raphson
iteration.

A similar non-linear eigenvalue problem was posed by Schandl \& Meyer
(1994).  In their model, the external torque is due to a coronal wind.
However, an earlier and incorrect form of the angular momentum
equation was adopted, and the actual solution of the eigenvalue
problem was not given.

\subsection{Numerical investigation}

The easiest way to find non-linear solutions is to set $r_{\rm b}$
close to a value corresponding to marginal stability of any mode, and
to search for a solution with a small value of $\beta$ and with
$\omega_{\rm p}$ close to the frequency of the marginal mode.  The
branches of solutions can then be followed quasi-continuously as
$r_{\rm b}$ is varied.

The results of this calculation are shown in Fig.~3, where we have
taken $\alpha=0.3$ and $\epsilon=0.1$.  The tidal torque is not
included.  We compare the results of a {\it simplified model}, in
which self-shadowing is neglected and linear hydrodynamics assumed,
with the {\it full model}.  The eigenvalue $\beta$, which is the
inclination of the inner disc, is plotted as an indication of the
`amplitude' of the solution.

\begin{figure*}
  \centerline{\epsfbox{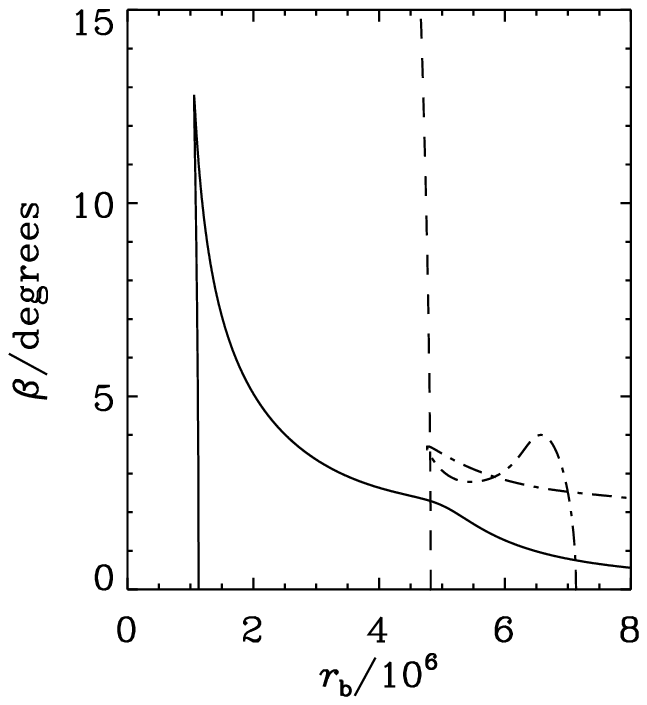}\qquad\epsfbox{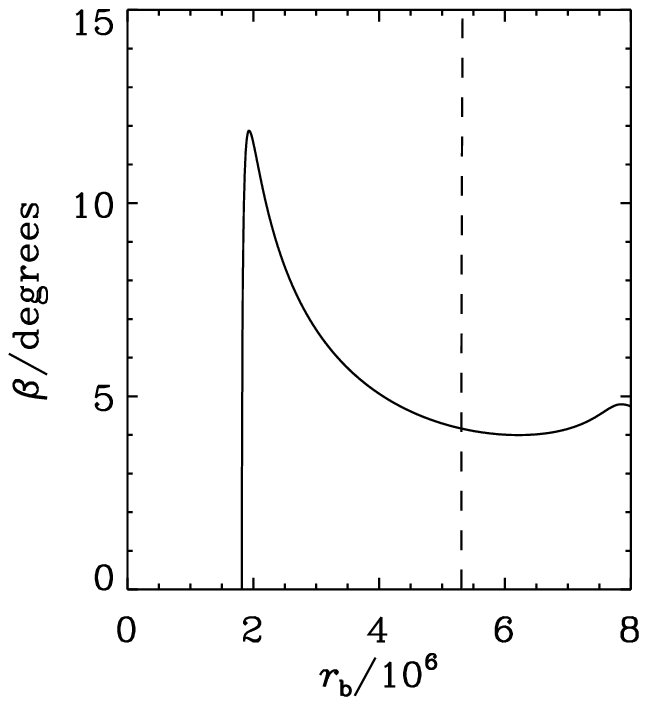}}
  \caption{Bifurcation diagram of non-linear solutions representing steadily
    precessing discs.  The inclination of the inner edge is plotted as
    an indication of the `amplitude' of the solution.  Left:
    simplified model.  Right: full model.  The various curves
    represent distinct branches of solutions produced at successive
    bifurcations of the flat solution (corresponding to the horizontal
    axis).  The solid line, branch 0, appears at the point of marginal
    stability, where mode 0 of the flat disc becomes unstable.  The
    dashed line, branch 1, extends to larger values of $\beta$ (not
    shown). Here $\alpha=0.3$ and $\epsilon=0.1$; the tidal torque is
    not included.}
\end{figure*}

In the simplified model, a branch of solutions appears at precisely
the point of marginal stability and rises very steeply out of a
subcritical bifurcation.  We label this branch $0$ because it results
from the instability of mode $0$ in the flat disc.  The branch reaches
a maximum value of $\beta$, turns around sharply and then extends over
a wide range of $r_{\rm b}$.  Further branches appear, as $r_{\rm b}$
is increased, at points where modes $1$, $2$, $\dots$ become unstable.
The branches do not interact with each other; it should be remembered
that $\beta$ is only a projection of an infinite-dimensional solution
space.

In the full model, the bifurcation diagram is qualitatively similar
except that the bifurcations are now supercritical.  The values of
$r_{\rm b}$ corresponding to marginal stability are greater, as noted
in Section~3.3.

In each case the solutions of branch 0 precess retrogradely, while
those of branch 1 precess progradely.  This shows that retrograde
precession can be obtained naturally even in the absence of a tidal
torque, as found by Wijers \& Pringle (1999) but contrary to the
conclusions of Maloney, Begelman \& Nowak (1998).  The precession
results partly from the radiation torque and partly from the internal
torque component with coefficient $Q_3$.  As explained by Wijers \&
Pringle (1999), when the mass input occurs well inside the outer
radius, the tilt usually increases outwards so that the radiation
torque causes retrograde precession.

\subsection{Stability of the steadily precessing discs}

Whether the solutions representing steadily precessing discs are
relevant to observed systems depends on their stability.  Theory
predicts that the first branch to bifurcate from the flat solution
(here, branch 0) will be initially stable if the primary bifurcation
is supercritical, but unstable if the bifurcation is subcritical.  All
subsequent branches are expected to be initially unstable.

In the subcritical case the first branch to bifurcate from the flat
solution is expected to regain stability at the turning point
(saddle-node bifurcation) where the minimum value of $r_{\rm b}$ is
achieved.  At some point along the branch, however, a secondary
bifurcation may occur at which the solution loses stability to a more
complex solution.  These expectations are shown schematically in the
insets of Figs~4--5.

We have attempted to verify these expectations by determining the
stability of the steadily precessing discs.  We begin by rewriting the
basic equations in a frame of reference that rotates at the precession
rate of the disc, and then linearize about the steady solution.  The
linearized equations admit normal-mode solutions proportional to
$\exp({\rm i}\omega t)$ and we seek to determine whether any solution
exists with ${\rm Im}(\omega)<0$, indicating the instability of the
solution.  This method is a generalization of the analysis of
Section~3 and the solutions obtained there can be recovered with the
more general method.

Owing to the technical difficulty of this calculation we do not
describe it in detail here.  In short, the results of the calculation
agree precisely with our expectations: in the simplified model where
the transition is subcritical, branch 0 is initially unstable, regains
stability at the turning point and then loses stability at a secondary
bifurcation; in the full model, the transition is supercritical and
branch 0 is initially stable before losing stability at a secondary
bifurcation. Branch 1 is unstable.

We have performed this calculation under two different assumptions
concerning the luminosity of the central source.  In the first case
the luminosity is constant and given by $L_\star=\epsilon\dot Mc^2$,
while in the second case the luminosity is variable and given by
\begin{equation}
  L_\star=\epsilon(-2\pi rv\Sigma)\bigg|_{r=r_{\rm i}}c^2.
\end{equation}
This more realistic prescription follows the suggestion of Wijers \&
Pringle (1999) that the luminosity should reflect the instantaneous
mass accretion rate at the inner radius, rather than the steady rate
supplied to the disc. Note that the distinction between constant and
variable luminosity does not affect any of our other results.
  
For the simplified model and in the case of constant luminosity,
branch 0 is stable in the interval $1.06\la r_{\rm b}/10^6\la2.45$.
With variable luminosity, it is stable for $1.06\la r_{\rm
  b}/10^6\la1.38$ (Fig.~4).  Therefore, in accordance with the
expectations of Wijers \& Pringle (1999), the variable luminosity
decreases the stability of the solution.  For the full model including
self-shadowing and non-linear hydrodynamics, the ranges of stability
are $1.82\la r_{\rm b}/10^6\la2.79$ and $1.82\la r_{\rm
  b}/10^6\la2.63$ (Fig.~5) for a constant and variable luminosity
respectively.

\begin{figure*}
  \centerline{\epsfbox{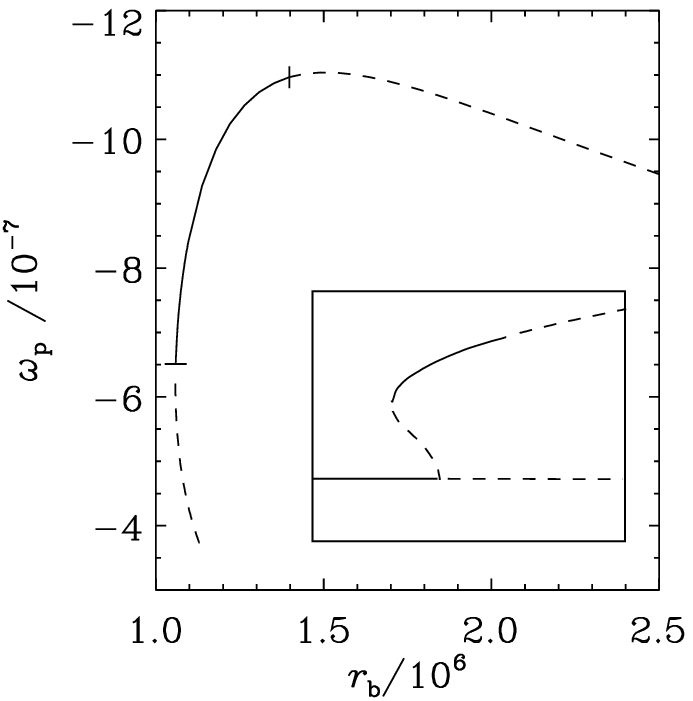}\qquad\epsfbox{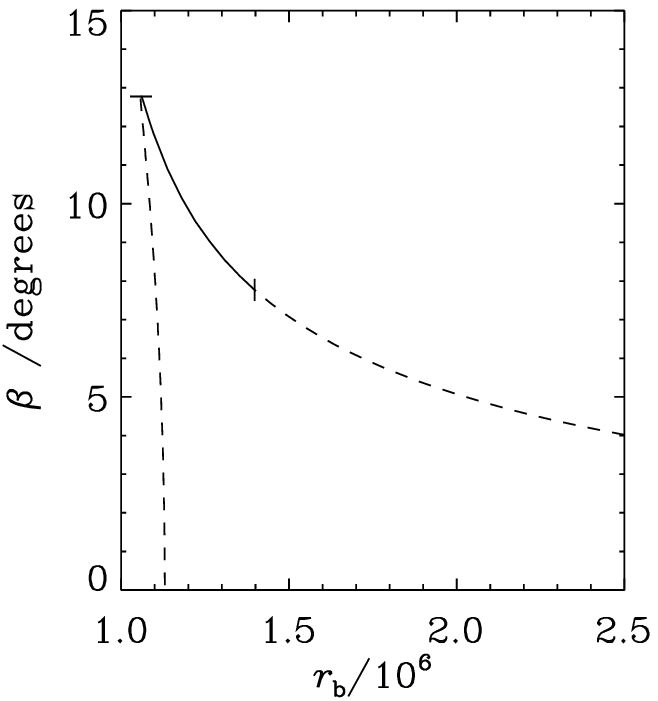}}
  \caption{Region of the bifurcation diagram where stable,
    steadily precessing solutions are found in the simplified model.
    These are shown with solid lines.  (Variable luminosity is
    assumed.)  Left: precession frequency in dimensionless units.
    Right: inclination of the inner edge.  Inset: schematic
    bifurcation diagram showing the expected stability properties of
    the non-linear solutions in the subcritical case.  Solid and
    dashed lines denote stable and unstable solutions, respectively.}
\end{figure*}

\begin{figure*}
  \centerline{\epsfbox{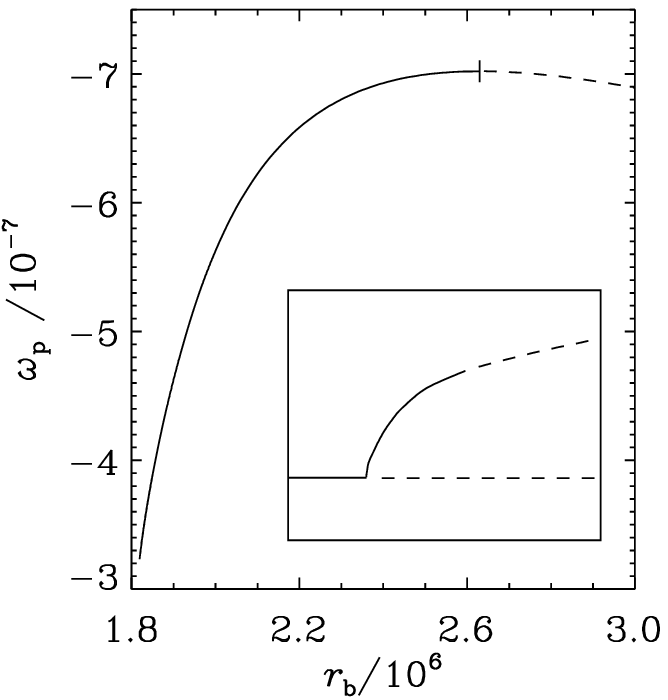}\qquad\epsfbox{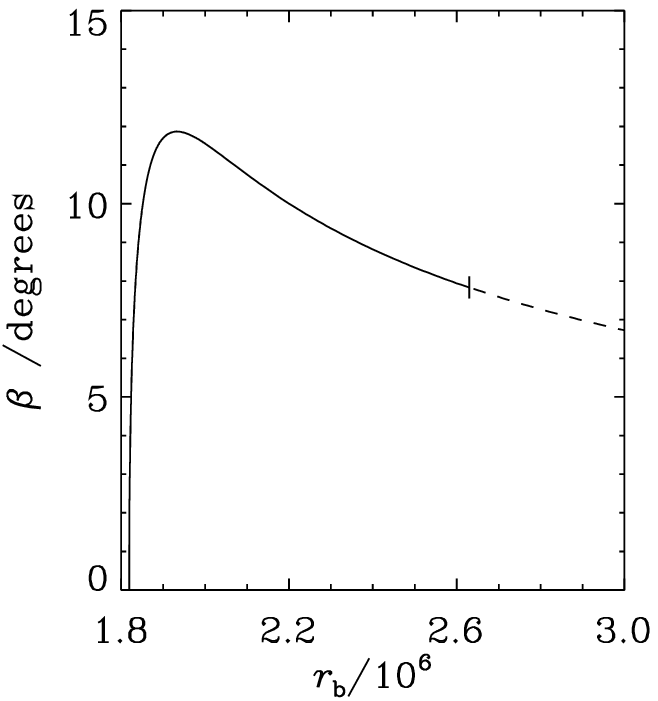}}
  \caption{Same as Fig. 4 but for the full model.  Inset: schematic
    bifurcation diagram showing the expected stability properties for
    the supercritical case.}
\end{figure*}

The interval of stability is small, but non-negligible, corresponding
to a variation of about 50\% in the control parameter, and this is
consistent with the results of Wijers \& Pringle (1999).  If robust,
this implies that strictly steadily precessing discs should be
relatively uncommon among X-ray binaries, and this is consistent with
observations.  The more complex solutions produced at the secondary
bifurcation would be quasi-periodic, i.e.  periodic in a rotating
frame of reference.  The accretion rate on to the compact object would
vary periodically in such a solution. Further bifurcations probably
lead to quasi-periodic or chaotic solutions which could be found only
by integrating the time-dependent equations.

\subsection{Properties of the stable non-linear solutions}

The solutions on the stable part of branch 0 (in the full model) have
a consistent and physically reasonable form.  As the control parameter
$r_{\rm b}$ is increased from the point of marginal stability, the
solution rapidly acquires a tilted and warped shape.  At $r_{\rm
  b}=2\times10^6$ (see Fig.~6) the inner and outer parts of the disc
have inclinations of $12\degr$ and $37\degr$, respectively, to the
binary plane.  As $r_{\rm b}$ is increased further, the outer tilt
remains close to $40\degr$ while the inner tilt decreases somewhat, as
seen in Fig.~5.

For a given solution, the tilt angle $\beta$ is an increasing function
of radius while the twist angle $\gamma$ (see equation
\ref{betagamma}) increases from $0$ at the inner edge to a maximum of
not more than $50\degr$ at an intermediate radius before declining
somewhat towards the outer edge.  The extent of the shadow increases
outwards, covering about half of the outermost annulus.  The amplitude
$|\psi|$ of the warp reaches a maximum, typically $0.4$--$0.6$, at an
intermediate radius.  The values of $|\psi|$ attained fully justify
our taking into account the non-linear expressions for the
coefficients $Q_i$ and the radiation reduction factors $g_1$ and
$g_2$.  The inner part of the disc is very flat, however, and
therefore passive, justifying our neglect of the regimes dominated by
Thomson opacity and/or radiation pressure.

Further geometrical properties are discussed in Section~5.4 below.

\begin{figure}
  \centerline{\epsfbox{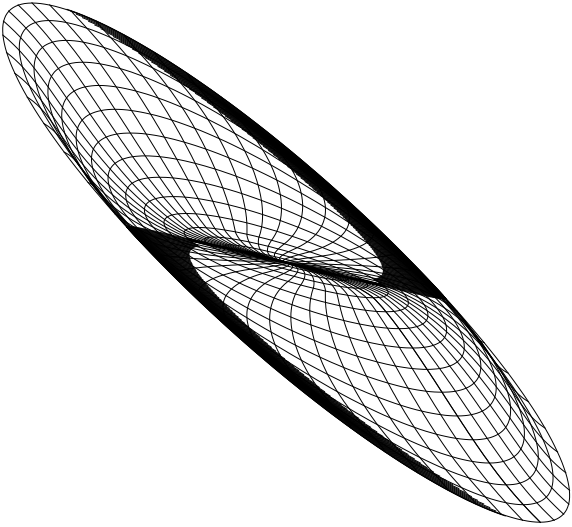}}
  \caption{Example of a steadily precessing disc on the stable part of
    branch 0 in the full model.  Here $\alpha=0.3$, $\epsilon=0.1$,
    $r_{\rm b}=2\times10^6$ and no tidal torque is included.  The
    precession is retrograde, with $\omega_{\rm p}=-5.6\times10^{-7}$.
    Lines of constant $r$ and $\phi$ (defined in Appendix~A) are
    plotted.  The binary system is viewed edge-on, showing the tilt
    and twist of the disc and the extent of the shadowed regions
    (black).  The inner and outer radii are inclined at $12\degr$ and
    $37\degr$, respectively, to the binary plane.}
\end{figure}

\section{Discussion}

\subsection{Stability of X-ray binaries to radiation-driven warping}

There has been some discussion as to whether X-ray binaries could or
could not be subject to radiation-driven warping.  We found that the
approximate criterion (\ref{pringle}) was close to the one found with
a more complete treatment including boundary conditions and
self-shadowing.  This criterion depends very strongly on the viscosity
parameter ($r_{\rm crit}\propto\alpha^{-4}$ for small $\alpha$) and on
the accretion efficiency ($r_{\rm crit}\propto\epsilon^{-2}$) as can
be seen from Fig.~1. In principle, it is always possible to find a
combination of $\alpha$ and $\epsilon$ for which a given system is
stable, slightly unstable, or highly unstable.

Estimating $\alpha$ in accretion discs is a long sought goal and the
best results so far have come from comparing the light-curves of dwarf
novae and soft X-ray transients to the predictions of the disc
instability model. According to these, $\alpha$ is in the range
0.1--0.3 when these systems are in outburst and unlikely to be larger
(e.g. Smak 1999; Cannizzo 1998).  This probably also holds for
persistent systems.  Unfortunately the value of $\alpha$ cannot yet be
reliably determined from numerical simulations of magnetohydrodynamic
turbulence in accretion discs.  While some local simulations suggest
values of $0.01$ or less, there are important dependences on the mean
magnetic field strength and on numerical details such as resolution
and boundary conditions (Brandenburg 1998).  More recent, global
simulations of cylindrical discs or thicker tori suggests values of
$0.1$ or greater (Hawley 2000).  As for $\epsilon$, not all X-ray
binaries host maximally spun-up black holes for which a maximum
efficiency of 0.30 is reached (Thorne 1974; although see Gammie
1999). For most systems $\alpha=0.3$ and $\epsilon=0.1$ are
therefore plausible, if somewhat optimistic.

\begin{figure}
  \centerline{\epsfbox{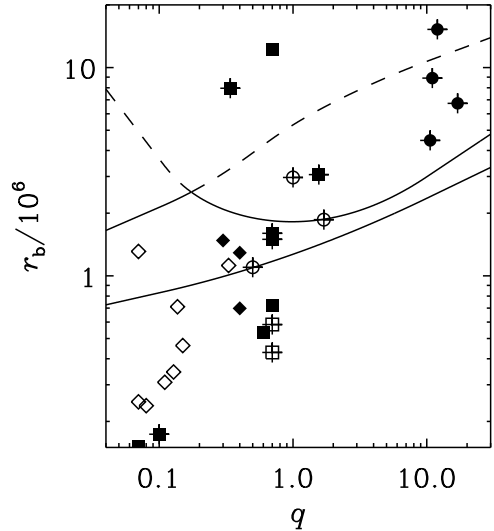}}
  \caption{Stability of X-ray binaries to radiation-driven warping
    ($\alpha=0.3$ and $\epsilon=0.1$). $r_{\rm b}$ is the binary
    separation in units of $GM_1/c^2$ and $q$ is the mass ratio
    $M_2/M_1$.  The upper two lines correspond to the first two
    bending modes when $r_{\rm add}$=$r_{\rm c}$. For $q>0.2$, mode 0
    is unstable first (solid line) and mode 1 is shown as a dashed
    line.  For $q<0.2$, the situation is reversed.  The lower solid
    curve corresponds to mode 1 for $r_{\rm add}$=$r_{\rm o}$.  The
    critical $r_{\rm b}$ for mode 0 is then much higher and is not
    plotted.  Also shown are the observed properties of a sample of
    systems (see Table~1). High-mass X-ray binaries are shown with
    circles, soft X-ray transients with diamonds, other low-mass X-ray
    binaries as squares. Black hole candidates appear as open symbols
    and neutron star primaries as solid symbols. Crosses represent
    systems for which a long time-scale has been reported.}
\end{figure}

\subsubsection{Dependence on the mass ratio $q$}

We now examine the dependence of the stability criterion on the mass
input radius $r_{\rm add}$ and on the mass ratio $q=M_2/M_1$ with
$\alpha=0.3$, $\epsilon=0.1$ and no tidal torque.  As in Sections~3
and~4, we first assume that the mass input radius $r_{\rm add}$ is the
circularization radius $r_{\rm c}$.  The ratios $r_{\rm c}/r_{\rm b}$
and $r_{\rm o}/r_{\rm b}$ are interpolated as functions of $q$ from
the tables of Lubow \& Shu (1975) and Papaloizou \& Pringle (1977).
The variation of the critical radius for the first two bending modes
is shown by the upper two lines in Fig.~7.  For $q\ga1$, $r_{\rm
  c}/r_{\rm o}$ is almost constant and the critical {\it disc} radius
varies only slightly.  The critical binary radius for both modes
therefore increases with increasing $q$ as $r_{\rm b}/r_{\rm o}$. For
$q\la1$, $r_{\rm c}/r_{\rm o}$ increases and the critical radius for
mode 0 becomes larger than what the approximate formula
(\ref{pringle}) predicts. For $q\la0.2$, the prograde mode 1 becomes
unstable first, presumably generating a branch of stable, progradely
precessing solutions.

The assumption that $r_{\rm add}=r_{\rm c}$ may not be realistic when
studying the stability of an initially flat disc. In this case, the
incoming stream from the companion will impact the disc at the outer
radius $r_{\rm o}$, although some of the matter may overflow to be
incorporated further inside (e.g. Armitage \& Livio 1998). When
$r_{\rm add}=r_{\rm o}$, mode 0 becomes unstable only for very large
$r_{\rm b}$ and mode 1 is the first to become unstable. Furthermore,
the critical radius at which mode 1 is unstable is smaller than when
mass is added at the circularization radius (Fig.~7).  This
calculation does not take into account the fact that the mass added at
$r_{\rm o}$ has too little angular momentum for a circular orbit at
that radius.  This would produce an extra torque which might affect
the stability properties.  If a disc can indeed be unstable to
radiation-driven warping when mass is added at the outer radius, but
stable when it is added the circularization radius, then such a system
might display warping cycles: an initially flat disc with mass input
at its outer edge becomes unstable and tilts; mass input then moves
towards $r_{\rm c}$ where the disc then becomes stable to warping and
resumes its initially flat shape. We refer to this region in Fig.~7 as
the `indeterminate instability zone'.  The opposite situation
was discussed by Wijers \& Pringle (1999) who argued that, because of
the higher surface density, a disc with mass input at the outer edge
would be more stable.  Our results suggest otherwise but a more
accurate treatment of mass input would be needed to reach a definite
conclusion.

\subsubsection{Comparison with observed systems}

\begin{table}
  \centering
  \caption{Parameters for the sample of X-ray binaries displayed in
Fig.~7.  The high-mass X-ray binaries appear first, followed by
low-mass X-ray binaries (Her X-1) and soft X-ray transients (GRO
J1655-40). Where there was no available estimate in the literature,
the value we assumed is shown in brackets. We took $M_1=1.4M_\odot$
for neutron star primaries. Periods are in days, $q=M_2/M_1$, $M_1$ is
in $M_\odot$ and the binary separation $r_{\rm b}$ is in units of $GM_1/c^2$.}
  \begin{tabular}{@{}lccccc@{}}
   System & $P_{\rm orb}$ & $P_{\rm long}$ & $q$ & $M_1$ & $r_{\rm b}/10^6$\\
\\
     Cen X-3 &   2.090 & 140 &  17.0 &  1.4 & 6.7\\
   4U1907+09 &   8.380 &  42 &  12.0 &  1.4 & 15.3\\
     SMC X-1 &   3.890 &  60 &  11.0 &  1.4 & 8.9\\
     LMC X-4 &   1.408 &  30 &  10.6 &  1.4 & 4.5\\
     Cyg X-1 &   5.600 & 142 &  1.70 &  10  & 1.9\\
     LMC X-3 &   1.706 &  99 &  0.50 &  5.0 & 1.1\\
      SS 433 &   13.10 & 164 & [1.0] &  [10] & 3.0\\
\\
     Her X-1 &   1.700 &  35 &  1.56 &  1.4 & 3.1\\
     Cyg X-2 &   9.844 &  78 &  0.34 &  1.4 & 8.0\\
     Sco X-1 &   0.788 &  62 & [0.7] &  1.4 & 1.6\\
       AC211 &   0.713 &  37 & [0.7] &  1.4 & 1.5\\
    GX 339-4 &   0.620 & 240 & [0.7] &  [5] & 0.6\\
   4U1957+11 &   0.390 & 117 & [0.7] &  [5] & 0.4\\
   4U1916-05 &   0.035 & 199 & [0.1] &  1.4 & 0.2\\
   4U1820-30 &   0.008 & 176 & [0.1] &  1.4 & 0.1\\
     Cir X-1 &   16.55 & --  & [0.7] &  1.4 & 12.2\\
    1746-370 &   0.238 & --  & [0.7] &  1.4 & 0.7\\
    1636-536 &   0.158 & --  & [0.6] &  1.4 & 0.5\\
   4U1626-67 &   0.029 & --  & [0.07] &  1.4 & 0.2\\
\\
GRO J1655-40 &   2.620 & --  &  0.33 &  7.0 & 1.1\\
     Aql X-1 &   0.800 & --  &  0.30 &  1.4 & 1.5\\
 Nova Oph 77 &   0.521 & --  &  0.15 &  4.9 & 0.5\\
 Nova Vel 93 &   0.285 & --  &  0.14 &  1.4 & 0.7\\
 Nova Mus 91 &   0.433 & --  &  0.13 &  6.2 & 0.3\\
GRO J0422+32 &   0.212 & --  &  0.11 &  3.6 & 0.3\\
   GS2000+25 &   0.344 & --  &  0.08 &  8.5 & 0.2\\
    A0620-00 &   0.323 & --  &  0.07 &  7.4 & 0.2\\
    V404 Cyg &   6.460 & --  &  0.07 & 12.3 & 1.3\\
     Cen X-4 &   0.629 & --  & [0.4] &  1.4 & 1.3\\
XTE J2123-05 &   0.250 & --  & [0.4] &  1.4 & 0.7\\
\end{tabular}
\end{table}

We have gathered data on a sample of high mass X-ray binaries showing
`super-orbital' modulation and on a sample of low-mass X-ray binaries
including soft X-ray transients\footnote{Radiation-driven warping can
  be important for soft X-ray transients only during their outbursts.
  They are then close to steady-state with $\dot{M}(r) \approx$
  constant so that we feel justified in including them here.} (Table
1). All of the systems listed by Wijers \& Pringle (1999) can be found
here.  The variability of the high-mass X-ray binaries in the sample
is not due to periodic accretion in an eccentric orbit.  There is
enough evidence that these are disc-fed and have circular orbits (e.g.
Bildsten et~al. 1997; Ilovaisky 1984; Petterson 1978).  They are all
(except LMC X-3, see Section~5.3.2) above the critical binary
separation for our chosen parameter values, due to the combination of
their long periods and high $q$.  Most of the low-mass X-ray binaries
are below or, depending on the assumption for the mass input radius,
slightly above the critical binary separation because of their shorter
orbital periods.  These conclusions cannot be affected by the errors
on the values of $q$ and $M_1$: $r_{\rm b}$ is weakly dependent on $q$
and $M_1$ for a low-mass X-ray binary of known orbital period so that
the values we used would have to dramatically overestimate $M_1$ {\em
  and} underestimate $q$ to make all the systems unstable.  This is
not likely.

An accurate treatment of mass input will probably yield a critical
binary separation in between the two extreme cases we have considered
and so should not modify qualitatively the above conclusions. Choosing
a higher value of $\alpha$ and/or $\epsilon$ would rapidly make
low-mass X-ray binaries more prone to radiation-driven warping but, as
argued above, there is nothing favouring higher values than
$\alpha=0.3$ and $\epsilon=0.1$.  In addition, a more realistic model
would take into account the tidal torque: although its effects are
stronger when $q$ is high (equation \ref{ftide}), Fig.~2 shows that
the tidal torque significantly increases the critical binary
separation (this is also found in the calculations of Wijers \&
Pringle 1999).  Furthermore, the comparison with steadily precessing
systems disfavours a smaller critical binary radius than shown here as
we shall see in the next section. We conclude that radiation-driven
warping may only be relevant to those low-mass X-ray binaries with
long orbital periods $\ga$ 1d (i.e. mostly soft X-ray transients).

\subsection{Steadily precessing systems}

In Section~4 we have investigated whether radiation-driven warping
could lead to a {\em stable} steadily precessing disc. We found such
solutions existed only in a narrow range of parameters.  This probably
explains why so few systems (Her X-1, SS 433, LMC X-4) show distinct,
repeatable cycles albeit with some variations.  There is ample
evidence that the cycles in these systems are not due to a variation
of the intrinsic luminosities (e.g. Margon 1984; Woo, Clark \& Levine
1995; Heemskerk \& van Paradijs 1989).

Most other `super-orbital' periods do not appear clearly on
periodograms and could be argued to be time-scales, modulations or
perhaps beating between different long-term periods. When steady
precession is possible, it is usually retrograde which is consistent
with the observations of Her X-1, SS 433 and LMC X-4 (Gerend \&
Boynton 1976; Margon 1984; Heemskerk \& van Paradijs 1989). For $q=1$,
we found the maximum binary separation for which steady behaviour
could exist was about 2.6 in units of $10^6$ $GM_1/c^2$. Her X-1 and
SS 433, which have $q\approx1$, lie very close to this value
considering the uncertainties: from table 1 we get $r_{\rm b}=3.06$
for Her X-1 and $r_{\rm b}=2.97$ for SS 433.  A slight {\em decrease}
of $\alpha$ or $\epsilon$ would provide better agreement, although
again we emphasize that the correct tidal torque for each system has
not been included.  It is also rewarding that, in Fig.~7, the next
closest source to the instability curve (for $r_{\rm add}=r_{\rm c}$)
happens to be LMC X-4.  All other systems are further away above the
curve or in the `indeterminate instability' region.

Because of the choice of units and the uncertainties in the different
parameters of the observed systems, it is difficult to evaluate the
precession period predicted by the model. In the range of stable
solutions for $q=1$, the precession frequency $\omega_{\rm p}$ varies
between 3.2$\times10^{-7}$ and 7.0$\times10^{-7}$ (Fig.~5) in units of
$C_\sI^{7/10}(\dot M/2\pi)^{3/10}(GM_1)^{-3/2}c^{5/2}$ (Section~2.3).
Using $C_\sI=4.4\times10^{12}$ in CGS units (Section~2.2.1), this
predicts precession periods for Her X-1 between 22d and 47d with $\dot
M\approx4\times10^{17}\,{\rm g}\,{\rm s}^{-1}$.  This estimate for
$\dot M$ comes from assuming $\epsilon=0.1$ and using the X-ray
luminosity quoted by Choi et~al.  (1994), corrected for the distance
estimate of Reynolds et~al.  (1997).  It also agrees with the analysis
of Cheng, Vrtilek \& Raymond (1995).  For Eddington-limited accretion
on to a $10M_\odot$ black hole with $\epsilon=0.1$ (i.e. $\dot
M\approx1.4\times10^{19}\,{\rm g}\,{\rm s}^{-1}$), we obtain
precession periods between 140d and 310d, consistent with the measured
period of 164d for SS 433. However, this system may be accreting at a
super-Eddington rate with low radiative efficiency. If $\epsilon$ is
significantly below $0.1$ then it is unlikely to be unstable to
radiation-driven warping. Despite the apparent agreement, we do not
feel warranted to make further quantitative comparisons because: (i)
the predicted precession period depends on the poorly known mass
accretion rate; (ii) the tidal torque almost certainly contributes to
the precession rate, and (iii) we have not fully investigated the
dependence of the precession period on the parameters of the model
($q$, $\alpha$, $\epsilon$ and $r_{\rm add}$).

Qualitatively, the dependence of the precession period on $M_1$ will
result in longer precession periods for black-hole candidates by about
an order of magnitude (e.g.  SS 433 against Her X-1 above and Corbet
\& Peele 1997)\footnote{This strengthens the interpretation of the
  106d period in the nuclear X-ray source of M33 as being due to disc
  precession around a 10 $M_\odot$ black hole (Dubus et~al. 1997).}.
Also, the frequency of mode 0 at marginal stability, where it
generates the branch of steadily precessing solutions, is only weakly
dependent on $q$ for $q\geq1$ (3.2--3.6$\times10^{-7}$ in
dimensionless units) and decreases to about $10^{-7}$ when $q=0.01$
where, in any case, very few systems are expected to show steady
behaviour (see previous section). Although a more detailed analysis
would be required, the range of steady precession frequencies will not
be much different when $q\geq1$.  This is consistent with the fact
that LMC X-4 has similar properties to Her X-1 ($P_{\rm orb}$, $P_{\rm
  long}$, $M_1$) except for $q\approx10$.

\subsection{Aperiodic variable systems}

A study of the behaviour of unstable systems would require a
time-dependent method and our aim here is to provide a basic framework
for interpretation. We have demonstrated in the previous sections that
unstable systems close to the stability limit will undergo steady
precession. As a system moves along the branch of solutions away from
the stable region, it will evolve to more complex solutions, probably
showing quasi-periodic behaviour before evolving towards chaos
(Section~4.3).  The exploratory calculations of Wijers \& Pringle
(1999) have shown such a sequence of behaviour. Here, the position of
each system in Fig.~7 can be used to infer its behaviour.

\subsubsection{Warp-driven variability}

The systems that seem beyond the stable range for steady precession
are Cen X-3, SMC X-1, 4U1907+09, Cyg X-2 and Cir X-1. SMC X-1 shows
clear quasi-periodic cycles in the RXTE ASM data between 50 and 60
days that are consistent with increased absorption by a warped disc
(Wojdowski et~al. 1998). In the RXTE ASM, Cen X-3 shows a succession
of {\sc on-off} transitions that might be reproduced by a variable
warp (Iping \& Petterson 1990; Priedhorsky \& Terrell 1983).  Tsunemi,
Kitamoto \& Tamura (1996) could find no correlations between the pulse
period history and the luminosity of this disc-fed pulsar.  They
concluded that the observed luminosity probably did not reflect the
mass accretion rate, which is consistent if the variability results
from a radiation-driven warp.

Although in the unstable region because of its extremely long $P_{\rm
  orb}$=16.6d, Cir X-1 has no reported long-term periodicity.
Johnston, Fender \& Wu (1999) proposed the neutron star is in an
eccentric orbit which would make the present study irrelevant.  This
may also be the case for 4U1907+09 for which a non-zero eccentricity
is reported (Bildsten et~al. 1997).

For $q\la0.2$, mode 1 becomes unstable first, implying that systems
there would most likely show {\em prograde} behaviour.  Cyg X-2 is
close to this region and, indeed, the 78d variability would be
consistent with a warped disc precessing progradely (Wijers \& Pringle
1999; Wijnands et~al. 1996; see also the discussion below of Paul
et~al. 2000).

\subsubsection{Marginally unstable or stable systems showing variability}

Some systems lie in the `indeterminate instability' zone, amongst
which are LMC X-3, Sco X-1, AC211 and Cyg X-1.  The listed $P_{\rm
  long}$ for these are more likely to be time-scales than real
periodicities.  Recently, Wilms et~al. (2000) presented a long-term
X-ray and optical study of the black hole candidate LMC X-3.  The
long-term spectral changes are not associated with changes in the
column absorption as expected for a warped disc.  The variations are
easier to explain if the source of hard photons (the corona) is
varying due to e.g. changes in the accretion rate.  In Cyg X-1, which
is right on the critical radius in Fig.~7, there seems to be evidence
for {\em both} a warp (Brocksopp et~al. 1999) and changes in the
accretion rate or geometry (Nowak et~al. 1999b).

A modulation of the accretion rate through the disc can occur on
time-scales of 10-100 days when a viscously unstable disc is
irradiated by a constant X-ray flux (Dubus et~al., in preparation).
This would also explain the time-scales and observations of GX 339-4
and 4U1957+11 which lie well below the critical binary radius and,
possibly, the very special cases of the ultra-short $P_{\rm orb}$
systems 4U1916-05 and 4U1820-30 (which does not even appear in Fig.~7)
where disc precession can be completely ruled out. In this case
modulations in the disc accretion rate cause the observed variability
while in the systems well above $r_{\rm crit}$ it is quasi-periodic or
chaotic warping that causes the variability. Nowak \& Wilms (1999a)
and Hakala, Muhli \& Dubus (1999) have favoured a warped disc for
4U1957+11 but there is actually little observational evidence to
support this.

The recent study of Paul, Kitamoto \& Makino (2000) comparing the
long-term variability of Cyg X-2 and LMC X-3 is particularly
interesting in this context.  They argue that in both cases the
observed variability is actually a combination of different more or
less stable periodic components.  This is consistent with the idea
that Cyg X-2 could be switching between different branches of warped
solutions. Contrary to LMC X-3, there is evidence in Cyg X-2 that the
variations in luminosity are not due to changes in $\dot{M}$
(Kuulkers, van der Klis \& Vaughan 1996).  If these were to be due to
varying obscuration then one would predict a hardening of the spectrum
at low intensities as the softer photons are absorbed (or because the
hard photons come from a little obscured corona). Paul et~al. indeed
find that the (5-12 keV)/(3-5 keV) hardness ratio is anti-correlated
to the intensity.  But there is no correlation for the (3-5
keV)/(1.5-3 keV) ratio and they conclude this does not support
obscuration.  Further observations at softer energies may be required
to settle this.

\subsubsection{Soft X-ray transients}

We have argued above that systems in the `indeterminate instability'
zone could show some hysteresis-type behaviour, switching between
warped and flat discs.  Such behaviour may enhance variations in the
mass transfer rate through the disc.  This varying low-amplitude warp
would not necessarily show up strongly as varying absorption, except
for high-inclination systems.  The long-$P_{\rm orb}$ soft X-ray
transients which lie in the `indeterminate instability' zone (V404
Cyg, Aql X-1, GRO J1655-40, Cen X-4) could very well host this type of
behaviour.

The soft X-ray transients can very roughly be divided into two groups:
one, associated with short orbital periods, showing the classic
fast-rise exponential decay (FRED) light-curve (e.g. A0620-00,
GS2000+25, Nova Mus 91, GRO J0422+32) and the other, with longer
orbital periods, showing a wider variety of light-curves with plateau
phases or erratic decays (see Chen, Shrader \& Livio 1997).  The
FRED-type light-curve can be accommodated within the disc instability
model, which provides the framework for understanding the outbursts of
soft X-ray transients and dwarf novae, if X-ray irradiation and disc
evaporation in quiescence are taken into account (Dubus et~al., in
preparation).

Yet this model does not provide an explanation of the variety of
light-curves observed for systems with long $P_{\rm orb}$.  A
particularly interesting case is GRO J1655-40 where the long plateau
during outburst and the observed anti-correlation between the optical
and X-rays were explained by the combined effects of increased mass
transfer from the secondary and variable warping (Esin, Lasota \&
Hynes 2000; see also Kuulkers et~al. 2000, who show the dipping
behaviour of the source changed during the outburst). Our study hints
that long-period soft X-ray transients may be subject to such variable
warping and this could partly explain their unusual outburst
light-curves.  Because of the dependence of the critical radius on
$M_1$, neutron star transients are more likely to be unstable to
radiation-driven warping and they should more often show deviations
from the prototypical FRED-type outburst light-curves.

\subsection{Warping and irradiation heating}

The stable, steadily precessing discs in the full model (Section~4),
which includes self-shadowing, all show a similar structure as
discussed in Section~4.4.  The shape of the disc when $r_{\rm
  b}=2\times 10^6$ is shown from the viewpoint of the compact object
in Fig.~8.  The tilt of the outer disc is quite large ($\sim40\degr$)
but Shakura et~al. (1998) deduced similar values for the disc tilt in
Her X-1 from RXTE observations. For SS 433, the jet is inclined from
its rotation axis by about $20\degr$ (Hjellming \& Johnston 1981),
also comparable to the values we obtain for $\beta$ at the inner edge
(see Fig.~5).

\begin{figure}
  \centerline{\epsfbox{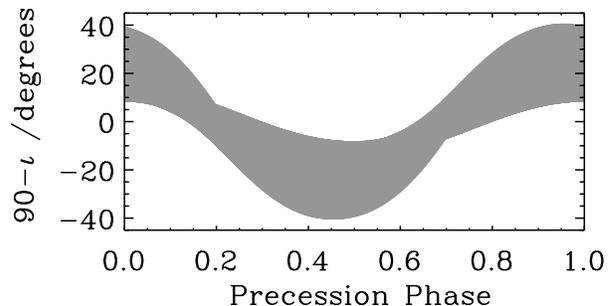}}
  \caption{The accretion disc for the solution in Fig.~6 as seen from
    the compact object.  The azimuth angle is equivalent to the
    precession phase. An observer at a high enough inclination $i$
    sees the disc (grey areas) periodically crossing his line of sight
    to the central source.}
\end{figure}

An observer at an inclination $i$ close enough to $90\degr$ will see a
succession of {\sc on-off} states as the central source disappears
behind the disc for certain phases. Observationally, the X-rays are
not fully eclipsed during the {\sc off} states, suggesting the
emission region is extended, very much like accretion disc corona
sources (Shakura et~al. 1998). Further studies would need to assess
whether such partial obscuration is compatible with the present model.
Interestingly, the disc in a system seen edge-on need not always
obscure the central source as can be seen from Fig.~8; but we would
need to include the effects of a non-zero disc thickness to verify
this.

Irradiation by the central source can heat the outer disc enough to
quench the thermal-viscous instability.  This is particularly
important in low-mass X-ray binaries where the standard disc
instability model without irradiation heating would predict many more
unstable systems than actually observed (van Paradijs 1996).
Irradiation is most important in the outer regions but these are
completely self-shadowed in a flat disc (Dubus et~al. 1999). If,
however, the disc is warped then the outer regions may see the central
source. Following Appendix~A, the amount of irradiation received by
each element in the disc is
\begin{equation}
  F_{\rm irr}=\bF\cdot\hatbn={\cal C}{L_{\star}\over{4\pi r^2}},
\end{equation}
where we have defined ${\cal C}=\be_r\cdot\hatbn$ for the warped disc
(Shakura \& Sunyaev 1973; Dubus et~al. 1999). 

We show in Fig.~9 the variation of ${\cal C}$ in the disc.  Dark areas
are able to intercept a large fraction of the flux from the central
source while white areas intercept little, or none if they are
shadowed.  The strongly irradiated regions form a two-armed spiral,
although it should be understood that the re-emitted radiation from
the two arms would be emitted on opposite sides of the disc.
The average value of ${\cal C}$ over an annulus, taking
self-shadowing into account, can be integrated like the radiation
torque and this is shown in Fig.~10.  In most of the outer region
${\cal C}$ is quite important ($\sim0.1$) and even a large albedo (the
fraction of the flux that is scattered) would not prevent significant
heating of the outer disc. For a solution very close to the stability
limit and thus corresponding to a much smaller tilt (about 2\degr), we
find ${\cal C}\sim 5\cdot10^{-3}$. Dubus et~al. (1999) found that
values of ${\cal C}$ of the order of $10^{-3}$ would be sufficient to
heat the disc significantly and stabilize it against the
thermal-viscous instability.  A warp can easily provide such values of
${\cal C}$ even with a significant albedo.  Taking the variations of
$H/r$ into account would probably not change these conclusions: $H/r$
would be expected to increase with radius, albeit with some variations
due to the warping (equation 43 of Ogilvie 2000), so that even a
partially shadowed outer disc would see enough of the X-ray flux.

\begin{figure}
  \centerline{\epsfbox{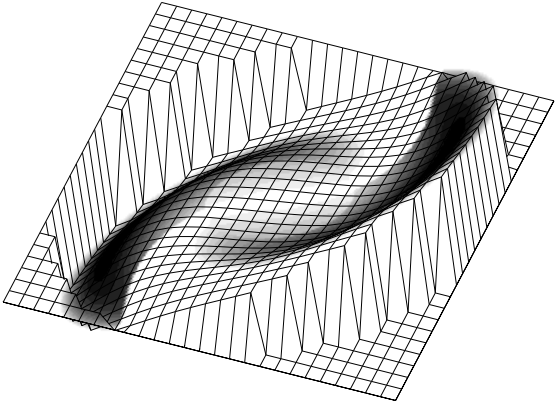}}
  \vskip0.6cm
  \caption{Another view of the stable, steadily precessing disc shown
    in Fig.~6.  The grey-scale shows variations of the coefficient
    $\cal C$.  The regions intercepting a large fraction of the flux
    are dark while those that intercept little or none (shadowed)
    appear white.  Re-emission from the two arms occurs on opposite
    sides of the disc.}
\end{figure}

\begin{figure}
  \centerline{\epsfbox{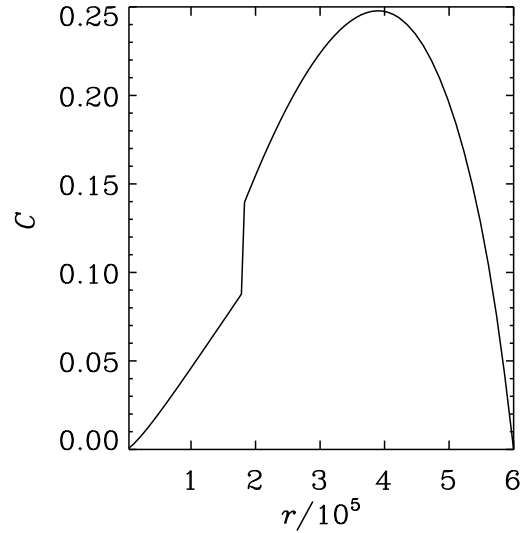}}
  \caption{Average value of $\cal C$ (see Fig.~9) as a function of $r$,
    taking self-shadowing into account.}
\end{figure}

While X-ray irradiation appears important for all low-mass X-ray
binaries (van Paradijs \& McClintock 1994), only a fraction would be
expected to be unstable to radiation-driven warping (Section~5.1). If
warping is to explain how the outer disc can intercept the X-ray flux
in all X-ray binaries then the warp must be produced differently (e.g.
winds, Schandl \& Meyer 1994; magnetic fields, although this may only
work close to the magnetosphere, Lai 1999).  The resulting disc will
probably also be able to intercept a large fraction of the flux.
Warping might have interesting consequences for Doppler tomography and
eclipse mapping: if irradiation heating dominates, the emission from
the disc is clearly asymmetric (Fig.~9; see also Still et~al. 1997,
where a search for such features is attempted for Her X-1); viscous
heating would also result in asymmetric emission since it is
significantly larger where the disc bends most (equation 44 of Ogilvie
2000).

\subsection{Uncertainties and neglected effects}

Our model rests on a number of assumptions and simplifications.  Some
of these may have affected our conclusions and we list them here as
suggestions for future investigations.
  
\begin{enumerate}
  
\item Regarding the hydrodynamics of the warped disc, the principal
  uncertainty is the modelling of the turbulent stress.  Although our
  results here favour a viscosity parameter $\alpha\ga0.1$, this
  depends on our assumption that the underlying effective viscous
  process is isotropic.  This difficult issue can be addressed only
  through numerical simulations of the turbulence (Torkelsson et~al.
  2000).
  
\item Although studying the dynamical effects of radiation, we have
  not taken into account the thermal effects of one-sided irradiation
  on the vertical structure of the disc. This has been
    considered by Ogilvie (2000) and could result in a fractional
    correction to equation (\ref{I}).
  
\item In our treatment of irradiation and self-shadowing, we neglected
  the non-zero thickness of the disc.
  
\item Our treatment of mass input to the warped disc is admittedly
  oversimplified (Section~2.2.2).
  
\item The correct inner boundary condition for different central
  objects is debatable (Section~2.2.3).

\end{enumerate}

\section{Conclusion}

We have studied the radiation-driven warping of accretion discs in
X-ray binaries.  The latest evolutionary equations were adopted, which
extend the classical alpha theory to time-dependent thin discs with
non-linear warps (Ogilvie 2000).  We have also developed accurate,
analytical expressions for the tidal torque and the radiation torque,
including self-shadowing, that can be easily implemented in numerical
calculations.

\begin{enumerate}
  
\item Using the complete set of equations, we re-examined the
  stability of discs to radiation-driven warping. We found that the
  critical binary separation is within an order of magnitude of the
  approximate criterion given by Pringle (1996) if the ratio of
  effective viscosities $\eta$ is given by equation (\ref{eta}).
  
\item Only the low-mass X-ray binaries with the longest orbital
  periods ($P_{\rm orb}\ga1$d) are likely to be unstable to
  radiation-driven warping.  This could explain the unusual outburst
  light-curves of the long-period soft X-ray transients.  The disc-fed
  high-mass X-ray binaries are more likely to be unstable.
  
\item We solved directly for several branches of non-linear solutions
  representing steadily precessing warped discs.  We studied the
  stability of these branches to further perturbations and found that
  typically only one branch is stable, and that only in a limited
  range of parameters. In this case, the precession is usually
  retrograde.
  
\item Discs that are beyond this range of parameters will probably
  show quasi-periodic or chaotic behaviour.
  
\item For $q\la0.2$, the radiation-driven instability may produce
  progradely precessing warped discs.
  
\item There may exist a certain range of parameters (the
  `indeterminate instability zone') for which a disc might cycle
  between being warped and being flat.
  
\item Our results are sensitive to assumptions concerning the
    effective viscosity of the disc.  If the turbulence acts on
    internal shearing motions similarly to an isotropic effective
    viscosity, then values of $\alpha$ exceeding $0.1$, and preferably
    closer to $0.3$, are required.

\end{enumerate}

Further studies should consider individual systems, including the
correct tidal torque on a case-by-case basis, and should make a more
detailed comparison with observations. For systems in which periodic
behaviour is not indicated, a time-dependent method should be used to
model their variability.

The present results confirm the conclusions of exploratory
calculations by Wijers \& Pringle (1999).  Our methods, which are
based on solving only ordinary differential equations, should be seen
as complementary to their time-dependent numerical approach.
High-precision solutions can be obtained rapidly using our approach,
and traced throughout the parameter space, but the non-linear dynamics
can only be followed up to the point at which the steadily precessing
discs become unstable to further perturbations.

We find that showing the binary separation as a function of the mass
ratio can be a powerful tool to discuss the behaviour of X-ray
binaries with respect to radiation-driven warping (Fig.~7). For the
systems selected, there is a very persuasive consistency between the
model and the observations: for instance, Her X-1, SS 433 and LMC X-4
are very well explained by stable steadily precessing discs. However,
most long-term variabilities observed in low-mass X-ray binaries
cannot be associated with a radiatively-driven warped disc and might
be due to modulations of the accretion rate. If warps are found to be
ubiquitous in low-mass X-ray binaries, then it is likely that other
driving mechanisms are at work.

\section*{Acknowledgments}

We thank Henk Spruit, Jim Pringle, Alastair Rucklidge, Jean-Pierre
Lasota, Rob Fender, Hannah Quaintrell and Mitch Begelman for
helpful discussions, and the University of Crete for hospitality when
this work was initiated. We acknowledge support from the European
Commission through the TMR network `Accretion on to Black Holes,
Compact Stars and Protostars' (contract number ERBFMRX-CT98-0195).
GIO is supported by Clare College, Cambridge.

\appendix

\section{Radiation torque}

In this section we derive an analytical expression for the radiation
torque density $\bT_{\rm rad}$.  Our discussion is based largely on
the analysis of Pringle (1996).

\subsection{Geometrical considerations}

Consider the disc to be composed of a sequence of concentric circular
rings.  The unit tilt vector $\bell(r,t)$ of the rings varies
continuously with radius $r$ and time $t$.  In the notation of Ogilvie
(1999),
\begin{equation}
  \bell=(\sin\beta\cos\gamma,\sin\beta\sin\gamma,\cos\beta),
  \label{betagamma}
\end{equation}
where $\beta(r,t)$ and $\gamma(r,t)$ are the Euler angles of the tilt
vector with respect to a fixed Cartesian coordinate system $(x,y,z)$.
Introduce the dimensionless complex variable
\begin{equation}
  \psi=|\psi|\,{\rm e}^{{\rm i}\chi}=
  r\left({{\partial\beta}\over{\partial r}}+
  {\rm i}{{\partial\gamma}\over{\partial r}}\sin\beta\right).
\end{equation}
Then
\begin{equation}
  |\psi|=\left|{{\partial\bell}\over{\partial\ln r}}\right|
\end{equation}
is a measure of the amplitude of the warp.  In particular,
\begin{equation}
  r{{\partial\bell}\over{\partial r}}=|\psi|
  \left[\matrix{\cos\beta\cos\gamma\cos\chi-\sin\gamma\sin\chi\cr
  \cos\beta\sin\gamma\cos\chi+\cos\gamma\sin\chi\cr
  -\sin\beta\cos\chi}\right],
\end{equation}
\begin{equation}
  r\bell\times{{\partial\bell}\over{\partial r}}=|\psi|
  \left[\matrix{-\cos\beta\cos\gamma\sin\chi-\sin\gamma\cos\chi\cr
  -\cos\beta\sin\gamma\sin\chi+\cos\gamma\cos\chi\cr
  \sin\beta\sin\chi}\right].
\end{equation}

The position vector of a point on the ring of radius $r$ is $\br=r\,\be_r$,
where (equation 7 of Ogilvie 1999, with $\theta=\pi/2$)
\begin{equation}
  \be_r=\left[\matrix{\cos\beta\cos\gamma\cos\phi-\sin\gamma\sin\phi\cr
  \cos\beta\sin\gamma\cos\phi+\cos\gamma\sin\phi\cr
  -\sin\beta\cos\phi}\right]
\end{equation}
is the radial unit vector and $\phi$ the azimuth on the
ring.\footnote{This definition of $\phi$ differs by $\pi/2$ from that
  of Pringle (1996).} Note that $\bell\cdot\be_r=0$, since points on
the ring lie in the plane that passes through the origin and is
orthogonal to $\bell$.  Now $(r,\phi)$ are (in general) non-orthogonal
coordinates on the surface of the disc.  The element of surface area
is
\begin{equation}
  {\rm d}\bS=\left({{\partial\br}\over{\partial r}}\,{\rm d}r\right)
  \times\left({{\partial\br}\over{\partial\phi}}\,{\rm d}\phi\right),
\end{equation}
which simplifies to
\begin{equation}
  {\rm d}\bS=\left[\bell+|\psi|\cos(\phi-\chi)\,\be_r\right]
  r\,{\rm d}r\,{\rm d}\phi.
\end{equation}
The unit normal to the surface is then
\begin{equation}
  \hatbn=\left[1+|\psi|^2\cos^2(\phi-\chi)\right]^{-1/2}
  \left[\bell+|\psi|\cos(\phi-\chi)\,\be_r\right].
\end{equation}

\subsection{Radiation}

Let the disc be illuminated by isotropic radiation from the central
source, of luminosity $L_\star$.  Then the energy flux is
\begin{equation}
  \bF={{L_\star}\over{4\pi r^2}}\,\be_r,
\end{equation}
except in regions of shadow.  The power incident on an illuminated
surface element ${\rm d}\bS$ is
\begin{equation}
  {\rm d}P=|\bF\cdot{\rm d}\bS|={{L_\star}\over{4\pi r^2}}
  |\psi||\cos(\phi-\chi)|\,r\,{\rm d}r\,{\rm d}\phi.
\end{equation}
Assume that the element absorbs the incident radiation and re-radiates
it immediately and uniformly from the same side.  Since the incident
radiation is parallel to the radius vector, the element receives no
torque from the absorption process.  However, it receives a reaction
force
\begin{equation}
  {\rm d}\bF=-{{2}\over{3}}{{{\rm d}P}\over{c}}\,(\pm\hatbn)
\end{equation}
from the re-radiation process, where $(\pm\hatbn)$ is the normal
pointing away from the disc on the side that received the radiation.
The correct sign is determined from the condition
$\be_r\cdot(\pm\hatbn)<0$, which implies
\begin{equation}
  {\rm d}\bF={{L_\star}\over{6\pi r^2c}}
  |\psi|\cos(\phi-\chi)\,\hatbn\,r\,{\rm d}r\,{\rm d}\phi.
\end{equation}
The corresponding torque is
\begin{equation}
  \br\times{\rm d}\bF={{L_\star}\over{6\pi rc}}
  q(1+q^2)^{-1/2}(\be_r\times\bell)\,r\,{\rm d}r\,{\rm d}\phi,
\end{equation}
where $q=|\psi|\cos(\phi-\chi)$.  The radiation torque density
$\bT_{\rm rad}(r,t)$ is defined such that $2\pi\bT_{\rm rad}\,r\,{\rm
  d}r$ is the torque acting on a ring of radius $r$ and of
infinitesimal radial extent ${\rm d}r$.  Therefore
\begin{equation}
  \bT_{\rm rad}=-{{L_\star}\over{12\pi rc}}\,\bell\times\ba,
  \label{trad}
\end{equation}
where
\begin{equation}
  \ba={{1}\over{\pi}}\int q(1+q^2)^{-1/2}\,\be_r\,{\rm d}\phi
\end{equation}
is a dimensionless integral.

Assume at this stage that there is no self-shadowing, so that the
integral is taken from $\phi=0$ to $\phi=2\pi$, or, equivalently, from
$\phi=\chi$ to $\phi=\chi+2\pi$.  Since $\be_r$ is a linear
combination of $\cos\phi$ and $\sin\phi$, the basic integrals required
to compute $\ba$ are
\begin{eqnarray}
  {{1}\over{\pi}}\int_0^{2\pi}(1+x^2\cos^2\alpha)^{-1/2}
  \cos^2\alpha\,{\rm d}\alpha&=&f(x),\nonumber\\
  {{1}\over{\pi}}\int_0^{2\pi}(1+x^2\cos^2\alpha)^{-1/2}
  \cos\alpha\sin\alpha\,{\rm d}\alpha&=&0,
\end{eqnarray}
where
\begin{equation}
  f(x)={{4}\over{\pi x^2(1+x^2)^{1/2}}}\left[(1+x^2)E(k)-K(k)\right].
\end{equation}
Here
\begin{eqnarray}
  K(k)&=&\int_0^{\pi/2}(1-k^2\sin^2\alpha)^{-1/2}\,{\rm d}\alpha,\nonumber\\
  E(k)&=&\int_0^{\pi/2}(1-k^2\sin^2\alpha)^{1/2}\,{\rm d}\alpha,
\end{eqnarray}
are the complete elliptic integrals of the first and second kinds,
respectively, of modulus $k=x/(1+x^2)^{1/2}$.
Thus
\begin{eqnarray}
  {{1}\over{\pi}}\int_0^{2\pi}q(1+q^2)^{-1/2}
  \cos\phi\,{\rm d}\phi&=&|\psi|\cos\chi\, f(|\psi|),\nonumber\\
  {{1}\over{\pi}}\int_0^{2\pi}q(1+q^2)^{-1/2}
  \sin\phi\,{\rm d}\phi&=&|\psi|\sin\chi\, f(|\psi|),
\end{eqnarray}
and so
\begin{equation}
  \ba=f(|\psi|)\,r{{\partial\bell}\over{\partial r}}.
\end{equation}
We obtain finally
\begin{equation}
  \bT_{\rm rad}=-{{L_\star}\over{12\pi c}}f(|\psi|)\,\bell\times
  {{\partial\bell}\over{\partial r}}.
\end{equation}
The factor $f(|\psi|)$ reduces the effectiveness of the radiation
torque when the disc is significantly warped.  It is a monotonically
decreasing function of its argument, with $f(0)=1$ and $f(x)\sim4/(\pi
x)$ as $x\to\infty$.  It has the Taylor series
\begin{equation}
  f(x)=1-{{3}\over{8}}x^2+{{15}\over{64}}x^4-{{175}\over{1024}}x^6+O(x^8).
\end{equation}

\subsection{Self-shadowing}

Consider the ring of radius $r$, with tilt vector $\bell$, and
consider an arbitrary ring of smaller radius $\hat r<r$, with tilt
vector $\hatbell\neq\bell$.  The smaller ring obscures the radiation
in the plane orthogonal to $\hatbell$ and casts shadows at two
diametrically opposite points on the larger ring.  These points lie in
the plane orthogonal to $\hatbell$ and are therefore defined by
$\br\cdot\hatbell=0$.  This determines their azimuth $\phi_{\rm s}$
according to
\begin{equation}
  \tan\phi_{\rm s}={{\hat\ell_x\cos\beta\cos\gamma+
  \hat\ell_y\cos\beta\sin\gamma-\hat\ell_z\sin\beta}
  \over{\hat\ell_x\sin\gamma-\hat\ell_y\cos\gamma}}.
\end{equation}
However, it is more convenient to measure the azimuth with respect to
the angle $\chi$, i.e. $\theta=\phi-\chi$.  We then find the covariant
expression
\begin{equation}
  \tan\theta_{\rm s}=-
  \left(\hatbell\cdot{{\partial\bell}\over{\partial r}}\right)\bigg/
  \left(\hatbell\cdot\bell\times{{\partial\bell}\over{\partial r}}\right).
  \label{thetas}
\end{equation}
Note that $\theta_{\rm s}$ is not uniquely defined since any multiple
of $\pi$ may be added to it.  Consider a single branch, therefore, and
follow $\theta_{\rm s}$ continuously as $\hat r$ varies from the inner
radius of the disc up to $r$.  Let $\theta_{\rm min}$ and $\theta_{\rm
  max}$ denote the minimum and maximum values obtained.  Then the
shadow on the ring considered consists of the two segments
$\theta_{\rm min}<\theta<\theta_{\rm max}$ and $\theta_{\rm
  min}+\pi<\theta<\theta_{\rm max}+\pi$.  If these overlap (i.e. if
$\theta_{\rm max}-\theta_{\rm min}>\pi$), the ring is completely
shadowed and the radiation torque is zero.  Otherwise, we must
reconsider the integral $\ba$ in the form
\begin{eqnarray}
  \ba&=&{{1}\over{\pi}}\int_{\theta_{\rm max}}^{\theta_{\rm min}+\pi}
  q(1+q^2)^{-1/2}\,\be_r\,{\rm d}\theta\nonumber\\
  &&+{{1}\over{\pi}}\int_{\theta_{\rm max}+\pi}^{\theta_{\rm min}+2\pi}
  q(1+q^2)^{-1/2}\,\be_r\,{\rm d}\theta\nonumber\\
  &=&{{2}\over{\pi}}\int_{\theta_{\rm max}}^{\theta_{\rm min}+\pi}
  q(1+q^2)^{-1/2}\,\be_r\,{\rm d}\theta.
\end{eqnarray}
Now the basic integrals required are
\begin{eqnarray}
  {{2}\over{\pi}}\int_0^\theta(1+x^2\cos^2\alpha)^{-1/2}\cos^2\alpha
  \,{\rm d}\alpha&=&g_1(\theta,x),\nonumber\\
  {{2}\over{\pi}}\int_0^\theta(1+x^2\cos^2\alpha)^{-1/2}
  \cos\alpha\sin\alpha\,{\rm d}\alpha&=&g_2(\theta,x),
\end{eqnarray}
where
\begin{eqnarray}
  g_1(\theta,x)&=&{{2}\over{\pi
  x^2(1+x^2)^{1/2}}}\left[(1+x^2)E(\theta,k)-F(\theta,k)\right],\nonumber\\
  g_2(\theta,x)&=&{{2}\over{\pi x^2}}\left[(1+x^2)^{1/2}-
  (1+x^2\cos^2\theta)^{1/2}\right].
\end{eqnarray}
Here
\begin{eqnarray}
  F(\theta,k)&=&\int_0^\theta(1-k^2\sin^2\alpha)^{-1/2}\,{\rm d}\alpha,
  \nonumber\\
  E(\theta,k)&=&\int_0^\theta(1-k^2\sin^2\alpha)^{1/2}\,{\rm d}\alpha,
\end{eqnarray}
are the incomplete elliptic integrals of the first and second kinds,
respectively, again of modulus $k=x/(1+x^2)^{1/2}$.
Thus
\begin{eqnarray}
  \ba&=&\left[g_1(\theta_{\rm min}+\pi,|\psi|)-
  g_1(\theta_{\rm max},|\psi|)\right]r{{\partial\bell}\over{\partial r}}
  \nonumber\\
  &&+\left[g_2(\theta_{\rm min}+\pi,|\psi|)-
  g_2(\theta_{\rm max},|\psi|)\right]
  r\bell\times{{\partial\bell}\over{\partial r}}.
\end{eqnarray}
Noting that $g_1(\theta+\pi,x)-g_1(\theta,x)=f(x)$ and
$g_2(\theta+\pi,x)-g_2(\theta,x)=0$, we confirm that the previous
result is recovered in the limit $\theta_{\rm max}-\theta_{\rm
  min}\to0$.

The elliptic integrals can be evaluated quickly and accurately using
Carlson's algorithms (see Press et~al. 1992).

\subsection{Scattering}

In reality, when X-rays from the central source are incident on the
disc, a significant fraction of the photons are scattered from the
surface rather than absorbed.  Wijers \& Pringle (1999) suggested that
this X-ray albedo might reduce the effectiveness of the radiation
torque.  However, we consider that the torque due to scattered
radiation will be very similar to that due to radiation that is
absorbed and re-emitted, because in each case the photons leaving the
disc have the same total energy flux as the incident radiation flux,
and the distribution of momenta of those photons is peaked around the
normal to the disc surface.  The non-isotropy of the Thomson
cross-section is quickly lost after a few scatterings and the photons
finally leaving the disc have paths almost perpendicular to the
surface.  The torque due to scattered radiation could only be much
smaller if the net effect of the scattering process was to divert the
momenta of the photons through a small angle, which would require
specular reflection from a slightly warped surface.  In conclusion,
the albedo of the disc is probably irrelevant to radiation-driven
warping.

\section{Tidal torque}

Assume that the companion star may be treated as a point mass $M_2$
describing a circular orbit in the $xy$-plane.  The position vector of
the companion star is then
\begin{equation}
  \br_2=r_{\rm b}(\cos\Omega_{\rm b}t,\sin\Omega_{\rm b}t,0),
\end{equation}
The time-averaged gravitational potential due to the companion star is
\begin{equation}
  \Phi_2(\br)=-{{GM_2}\over{2\pi}}\int_0^{2\pi}{{1}\over{|\br-\br_2|}}
  \,{\rm d}(\Omega_{\rm b}t).
\end{equation}
For $|\br|<|\br_2|$ we obtain the expansion
\begin{eqnarray}
  \lefteqn{\Phi_2=-{{GM_2}\over{r_{\rm b}}}\left[1+
  {{(s^2-2z^2)}\over{4r_{\rm b}^2}}+
  {{3(3s^4-24s^2z^2+8z^4)}\over{64r_{\rm b}^4}}\right.}&\nonumber\\
  &&\left.+{{5(5s^6-90s^4z^2+120s^2z^4-16z^6)}\over
  {256r_{\rm b}^6}}+\cdots\right],
\end{eqnarray}
where $s^2=x^2+y^2$.

Consider a circular ring of radius $r$, centred on the origin and
tilted at an angle $\beta$ to the binary plane.  Owing to the axial
symmetry of $\Phi_2$, we may take the tilt vector to be
\begin{equation}
  \bell=(\sin\beta,0,\cos\beta)
\end{equation}
without loss of generality.  The position vector of a point on the
ring is then
\begin{equation}
  \br=r(\cos\beta\cos\phi,\sin\phi,-\sin\beta\cos\phi),
\end{equation}
where $\phi$ is the azimuth on the ring.  It is clear from the
symmetry of the situation that the net torque on the ring has only a
$y$-component.  Now the $y$-component of the torque per unit mass is
\begin{equation}
  t_y=-z{{\partial\Phi_2}\over{\partial x}}+
  x{{\partial\Phi_2}\over{\partial z}}.
\end{equation}
When averaged over $\phi$ this gives
\begin{eqnarray}
  \lefteqn{\langle t_y\rangle=-{{GM_2}\over{r_{\rm b}}}
  \left[{{3r^2}\over{8r_{\rm b}^2}}\sin2\beta+
  {{45r^4}\over{1024r_{\rm b}^4}}(2\sin2\beta+7\sin4\beta)
  \right.}&\nonumber\\
  &&\left.+{{525r^6}\over{65536r_{\rm b}^6}}
  (5\sin2\beta+12\sin4\beta+33\sin6\beta)+\cdots\right].
\end{eqnarray}
The tidal torque density $\bT_{\rm tide}$ is obtained by multiplying
by the surface density $\Sigma$.\footnote{See Ogilvie (1999) for a
  demonstration that the surface density of a warped disc is
  independent of $\phi$.} Noting that $\be_z\cdot\bell=\cos\beta$ and
$\be_z\times\bell=\sin\beta\,\be_y$, we may write $\bT_{\rm tide}$ in
the covariant form
\begin{eqnarray}
  \lefteqn{\bT_{\rm tide}=
  -{{3GM_2\Sigma r^2}\over{4r_{\rm b}^3}}(\be_z\cdot\bell)
  (\be_z\times\bell)}&\nonumber\\
  &&\times\left\{1+{{15}\over{32}}\left({{r}\over{r_{\rm
  b}}}\right)^2\left[7(\be_z\cdot\bell)^2-3\right]\right.\nonumber\\
  &&\left.+{{175}\over{512}}\left({{r}\over{r_{\rm
  b}}}\right)^4\left[33(\be_z\cdot\bell)^4-30(\be_z\cdot\bell)^2+5\right]+
  \cdots\right\}.
  \label{ttide}
\end{eqnarray}

This series converges quite rapidly.  The magnitude of the fourth term
(not shown) in the series in braces is less than $0.0027$ when
$r/r_{\rm b}<0.3$.  This suggests that the series as truncated above
is accurate to within a fraction of a percent for all the calculations
in this paper.  Of course, higher-order terms can easily be calculated
if desired.

If the torque is {\it not} averaged over the binary orbital period, it
is found to contain an additional, oscillatory component with
frequency $2\Omega_{\rm b}$.  The oscillatory torque causes a small
wobbling motion but is unlikely to be important otherwise, except in
very thick discs, or discs much smaller than the standard tidal
truncation radius (Lubow \& Ogilvie 2000).

\label{lastpage}

\end{document}